\documentclass{aa}

\usepackage{color}
\usepackage{graphicx}
\usepackage{rotating}
\usepackage{longtable}

\def   \araa {{\rm {ARA\&A}}}
\def   \apj {{\rm {ApJ}}}

\def   \apjs {{\rm {ApJS}}}

\def   \aap {{\rm {A\&A}}}

\def   \mnras {{\rm {MNRAS}}}

\def   \apjl {{\rm {ApJL}}}

\def   \pasj {{\rm {PASJ}}}

\def   \ao   {{\rm {ApOpt}}}
\begin{document}

\title{Phase correction for ALMA - Investigating water vapour radiometer scaling:\\ The long-baseline science verification data case study}
\titlerunning{Investigating water vapour radiometer scaling}

\author{L.~T.~Maud\thanks{E--mail: maud@strw.leidenuniv.nl (LTM)}$^{,}$\thanks{{\sc python} code: http://www.alma-allegro.nl/wvr-and-phase-metrics/wvr-scaling/}\inst{1} \and R.~P.~J.~Tilanus\inst{1} \and T.~A.~van~Kempen\inst{2,1}  \and M.~R.~Hogerheijde\inst{1}  \and M.~Schmalzl\inst{1} \and I.~Yoon\inst{1,3} \and Y.~Contreras\inst{1} \and M.~C.~Toribio\inst{1} \and Y.~Asaki\inst{4,5} \and W.~R.~F.~Dent\inst{5} \and E.~Fomalont\inst{3,5} \and S.~Matsushita\inst{6}}

  \authorrunning{L.~T.~Maud et al.}
  
  \institute{Leiden Observatory, Leiden University, PO Box 9513, 2300 RA Leiden, The Netherlands
    \and SRON Netherlands Institute for Space Research, Sorbonnelaan 2, 3584 CA Utrecht
    \and National Radio Astronomy Observatory, 520 Edgemont Road, Charlottesville, VA 22911, USA
    \and National Astronomical Observatory of Japan (NAOJ) Chile Observatory, Alonso de Cordova 3107, Vitacura 763 0355, Santiago, Chile
    \and Joint ALMA Observatory (JAO), Vitacura 763 0355, Santiago, Chile
    \and Academia Sinica Institute of Astronomy and Astrophysics, PO Box 23-141, Taipei 10617, Taiwan, R.O.C }

\date{Received Day month year / Accepted day month year }




\abstract{The Atacama Large millimetre/submillimetre Array (ALMA) makes use of water vapour radiometers (WVR), which monitor the atmospheric water vapour line at 183 GHz along the line of sight above each antenna to correct for phase delays introduced by the wet component of the troposphere. The application of WVR derived phase corrections improve the image quality and facilitate successful observations in weather conditions that were classically marginal or poor. We present work to indicate that a scaling factor applied to the WVR solutions can act to further improve the phase stability and image quality of ALMA data. We find reduced phase noise statistics for 62 out of 75 datasets from the long-baseline science verification campaign after a WVR scaling factor is applied. The improvement of phase noise translates to an expected coherence improvement in 39 datasets. When imaging the bandpass source, we find 33 of the 39 datasets show an improvement in the signal-to-noise ratio (S/N) between a few to $\sim$30 percent. There are 23 datasets where the S/N of the science image is improved: 6 by $<$1\,\%, 11 between 1 and 5\,\%, and 6 above 5\,\%. The higher frequencies studied (band 6 and band 7) are those most improved, specifically datasets with low precipitable water vapour (PWV), $<$1\,mm, where the dominance of the wet component is reduced. Although these improvements are not profound, phase stability improvements via the WVR scaling factor come into play for the higher frequency ($>$450\,GHz) and long-baseline ($>$5\,km) observations. These inherently have poorer phase stability and are taken in low PWV ($<$1\,mm) conditions for which we find the scaling to be most effective. A promising explanation for the scaling factor is the mixing of dry and wet air components, although other origins are discussed. We have produced a {\sc python} code to allow ALMA users to undertake WVR scaling tests and make improvements to their data.}


\maketitle

\section{Introduction}
\label{intro}
Interferometric observations in the submillimetre/millimetre regime are strongly affected by the troposphere. Primarily, radiation is absorbed such that the transmission from an astronomical source is reduced and secondly, spatially and temporally variable delays in the path length of the source signal are introduced (i.e. refraction). The main components of the troposphere, i.e. oxygen, nitrogen dry, and water vapour wet, are the primary causes of the two phenomena. The signal absorption is irreversible and the lost signal strength cannot be recovered, however the variable delay in the path length of the source signal to each antenna in the interferometric array can be accounted for and (partially) corrected. In principle these effects are amenable to correction in the data processing stages \citep[e.g.][]{Hinder1971}. If one observes with baselines smaller than the characteristic length scale where the delays vary significantly and samples are faster than the temporal variation, imaging may be possible. In practice depending on instrumentation, capabilities, and calibrator availability, the observing strategies are adjusted such that imaging of astronomical sources can be accomplished with a reasonable accuracy.

\citet{Matsushita2017} have presented the first study of the atmospheric phase characteristics from the ALMA long-baseline campaign comprised of test data from 2012 to 2014. The 2014 campaign specifically focussed on baselines from 5 to 15\,km. The path length delays caused by the atmosphere, which are seen as phase fluctuations by an interferometer, increase with baseline length and follow a power-law slope of $\sim$0.6, although generally after $\sim$1$-$2\,km the slope becomes shallower to $\sim$0.2$-$0.3. After the application of phase corrections with the water vapour radiometer (WVR) system the phase fluctuations are decreased by more than half in many cases. However, there are still residual phase variations that remain unaccounted for, even when considering instrumental errors. We attempt to provide an extra step in reducing the phase variations via the application of a scaling factor in the WVR solutions.

\subsection*{Troposphere and its effects}
Generally the dry air components of the troposphere are well mixed and in near pressure equilibrium such that total column densities and pressures are slowly variable with time. However, the temperature of the dry air varies rapidly due to local heating and cooling, hydrostatic temperature variations, and wind-induced turbulence \citep{Nikolic2013}. Dry air only has a minor effect on the absorption of astronomical signals, but can have large refractive effects (in terms of variable delays in path length), often regarded as seeing. The delays are much more pronounced at the optical and infrared wavelengths and independent strategies such as adaptive optics (AO) \citep{Davies2012} have been employed to deal with distortions on $<$1 second timescales. The effects of dry air at submillimetre/millimetre wavelengths may still be measurable but have been thought to be small and vary on longer timescales \citep[e.g.][]{Hinder1972}. Interestingly, \citet{Matsushita2017} have found a few rare cases in which the phase variation with baseline length is particularly different to the generally understood atmospheric structure function, which could be due to predominantly dry air fluctuations.

The effects of the wet air, the variable water vapour cells, are the main cause of the refraction at submillimetre/millimetre wavelengths. The dipole moment of water makes water vapour, the wet component in the troposphere, a strong absorber at submillimetre/millimetre wavelengths and significantly increases the refractive index of the air. Because the water vapour is not well mixed there are localised pockets of air with different refractive indices. In what is called the `frozen-screen' hypothesis \citep{Taylor1938}, these pockets, or turbulent eddies, are assumed to be fixed in the atmospheric layer that advects over an interferometric array \citep{Thompson2017}. 
Thus, these cause various delays in the path length (variable in time and position) along the line of sight to each antenna. Interferometers are sensitive to the variations in path length, the interferometric phase difference, between pairs of antennas that form a baseline. For a given baseline (distance and orientation) the line-of-sight path to each component of an astronomical source has an intrinsic phase that relates the measured intensities to their location in an image. Thus any additional variable atmospheric delays that cause anomalous phase changes on many baselines making up an array have the effect of blurring the interferometric image; this is analogous to the effect of seeing at optical and infrared wavelengths. The introduced delays scale linearly with the difference in precipitable water vapour ($\Delta$PWV) between an antenna pair (excluding dispersive effects) and linearly with frequency. The correlated signals between pairs of antennas (the visibilities $V = V_oe^{i\phi}$) become partly decorrelated as a result of the phase noise. The reduced coherence for the visibilities is given by

\begin{equation}
  \label{eqn0}
\langle V \rangle = V_o \times \langle e^{i\phi} \rangle = V_o \times e^{-\phi^2_{rms}/2}
,\end{equation}

\noindent where $\phi_{rms}$ (in radians) is assumed to be Gaussian random phase noise occurring during the observations of the targeted source \citep{Thompson2017}. Phase errors also cause inaccuracies and incorrect features in synthesis images. Figure 5 of \citet{Carilli1999} illustrates the problem associated with image inaccuracies, such as changed source structures and anomalous features. 

In the case where one assumes the variations in the water vapour content are driven by fully developed turbulence, the Kolmogorov turbulence theory \citep{Coulman1990} is shown to predict that the RMS phase variations are a function of baseline length of the form

\begin{equation}
\label{eqn1}
\phi_{rms}(b) = \frac{K}{\lambda} b^{\alpha} (degrees)
,\end{equation}


\noindent where $b$ is baseline length, $\lambda$ is the observing wavelength, and $\alpha$ is the turbulent theory exponent. The parameter $K$ is related to atmospheric conditions and is typically around 100 at Chajnantor as found in early ALMA site tests compared to nearly 300 for the Very Large Array (VLA) \citep{Carilli1996}. The RMS phase variations rise with baseline length and are thought to continue up to an ‘outer length scale’ \citep{Carilli1999}. Theory predicts three components: a thick turbulent component (3D), where $\phi_{rms}\,\propto\,b^{0.83}$; a thin screen (2D), where $\phi_{rms}\,\propto\,b^{0.33}$ ; and on the largest length scales, where $\phi_{rms}$ is independent of $b$ ($\alpha$=0.83, 0.33, and 0, respectively). This is understood as the atmospheric spatial structure function (SSF), where a measure of phase noise is compared with the baseline length \citep[e.g.][]{Carilli1999,Matsushita2017}. The thick to thin screen transition is thought to occur when the baseline length is approximately equivalent to the scale height of the refractive atmospheric layer (i.e. thickness or vertical extent of the turbulent layer). Although the exponent ($\alpha$) decreases with increasing baseline length, the measurement of radiation from astronomical sources still becomes less accurate and less efficient for longer baselines. The sought after long-baseline ($>$5\,km) observations with ALMA that can attain the highest spatial resolution (e.g. tens of milli-arcseconds at 230\,GHz and $<$10\,mas $>$850\,GHz) will be most difficult owing to large phase fluctuations for $>$5\,km length baselines. An outer length scale could be a saving grace for such observations, where the amplitude of the phase fluctuations would become independent of baseline length. However, recent results have indicated the continual increase of phase fluctuations at ALMA out to $\sim$10$-$15\,km baselines \citep{ALMA2015a,Matsushita2017}.

\subsection*{Counteracting the atmosphere}
Without correction there are severe constraints on the maximum time and maximum baseline lengths that can be used in observations. Excluding only the very shortest baselines, over a short timescale of the order of tens of seconds to minutes the phase RMS would cause complete decorrelation for observations \citep{Matsushita2017}. Thus, various strategies can be used to counter the effect of atmospheric phase fluctuations. These techniques are: self-calibration, phase referencing, paired antennas, and water vapour radiometry.

In short, self-calibration requires a sufficiently strong source with a high surface brightness such that it can itself be used to calibrate the differential phase delays (for more details see \citealt{Pearson1984} and \citealt{Cornwell1999}).

Phase referencing is the standard practice for most interferometric observations. Here a point source calibrator is observed interspersed with the observations of the astronomical target(s) at regular intervals. In the data processing the phase solutions are transferred and interpolated from the calibrator to the source. The phase variations occurring on timescales longer than the referencing are now corrected, although those occurring when observing the science target are not.

\citet{Asaki1996} describe the paired antenna method. Here a sub-array of antennas is used to permanently observe a calibrator while the remaining are used to observe the target simultaneously. The solutions are again transferred in the data processing stages, allowing a calibration in `real-time' at the expense of fewer antennas observing the science target and slightly different lines of sight.

For a more detail overview of these techniques, see \citet{Carilli1999}.

\subsection*{Water vapour radiometry}
It was realised in the 1960s that radiometers could be used for sensing the water vapour content, and perhaps to actively correct for the phase errors for millimetre interferometers \citep{Baars1967, Barrett1962}. Monitoring the changes in the PWV along the line of sight of each antenna can be used to correct phase fluctuations by calculating the differential delays caused. Most mm-regime interferometers have trialled or implemented WVR systems, usually based on the 22\,GHz water transition (IRAM Plateau de Bure Interferometer, PdBI, \citealt{Bremer2002}; Very Large Array, VLA, \citealt{Butler2000}; Owens Valley Radio Observatory, OVRO, \citealt{Woody2000}; Australia Telescope Compact Array, ATCA, \citealt{Indermuehle2013}). The system operated at ALMA is based upon the 183\,GHz water transition and was initially tested on a baseline between the Caltech Submillimeter Observatory and the James Clerk Maxwell Telescope \citep{Wiedner2001}. 
The PWV at very dry, i.e. high altitude or arctic, sites can become so low that measurements using the 22\,GHz line become relatively insensitive.
Although the 183\,GHz line is easier to saturate this rarely occurs given the typically low PWV at the ALMA site. The full development of the WVR system for ALMA is detailed in a series of papers and ALMA memos from the mid-2000s \citep[e.g.][]{Delgado2000,Nikolic2007,Nikolic2012, Nikolic2013,Stirling2005}. Specifically the final WVR correction has been showcased in \citet{Nikolic2013}.

ALMA observations are always delivered with the WVR data that is used to calculate the differential phase solutions in the {\sc wvrgcal} code \citep{Nikolic2012}. When applied these can significantly correct the phases of the data. The \citet{ALMA2015a} have reported that in general around half of the short-term phase fluctuations are removed and therefore the proportion of time that phase referenced observations can be used to make good images is increased. The improvements after WVR application are relatively lower for dryer conditions although there is an option, which was untested prior to our work, to scale the WVR solutions and possibly improve the corrections \citep{Nikolic2013}. \citet{Matsushita2017} have indicated that usually the improvement ratios for the RMS phase are $\sim$1.7 for conditions under 1\,mm PWV and around $\sim$2.4 for conditions above, averaged over all baselines. These authors have also found that there are still residual phase fluctuations that remain and these are larger than the specification for ALMA after accounting for instrumental factors. The \citet{ALMA2015a} report states that in conditions where PWV$<$2\,mm and with clear skies the RMS phase should be as low as $\sim$20\,$\mu$m, although this is not the case.

This paper presents the results of using an empirically established scaling factor to scale the WVR solutions that are applied to the data. The scale factor is introduced in the {\sc wvrgcal} code (see \citealt{Nikolic2013}) and we show that it can improve the phase statistics over the standard WVR correction in preferentially dryer atmospheric conditions. Improving the phases by means of WVR scaling has the effect of reducing losses caused by decoherence during the observations on any source, including the science target for which there is generally no other means of correcting these phases (except where self-calibration is possible). Improved coherence can result in, for some cases, a higher dynamic range and potentially a higher fidelity of the science target image. An improved WVR calibration implies that observations could also take place in classically worse conditions compared to when the scaling factor is not applied.

\section{Observations and reduction}
\label{obs}
We conduct our WVR scaling investigation on the publicly available ALMA Science Verification (SV) observations taken during the long-baseline campaign. These include the observations of the asteroid Juno at band 6 (2011.0.00013.SV), the well-studied asymtotic giant branch (AGB) star Mira at bands 3 and 6 (2011.0.00014.SV), the young protostar with circumstellar disk, HL Tau, at bands 3, 6, and 7 (2011.0.00015.SV) and the lensed ultra-luminous starburst galaxy SDP.81 at bands 4, 6, and 7 (2011.0.00016.SV). For a detailed description of the specific spectral window settings of the data, see the ALMA SV page\footnote{https://almascience.nrao.edu/alma-data/science-verification} as some observations contain a mix of wide and narrow bandwidths as both continuum (time division mode; TDM) and spectral line (frequency division mode; FDM) modes were used. 

Each individual ALMA dataset as part of these SV datasets, i.e. an execution block (EB), has a typical observing time of $\sim$1\,hour. For some datasets, e.g. HL Tau and SDP.81, each EB was scheduled with different start times by one to a few hours to obtain good $(u,v)$ coverage (i.e. visibility coverage) using the aperture synthesis technique \citep{Thompson2017} required to image the target accurately given the long-baseline test array configuration; repeating each EB at the same time of day, for the same source elevation, would mean some (u,v) coordinates would have been under- or poorly sampled. Each EB has a specific unique identification ID (UID) of which we only refer to the suffix as a means to identify the various datasets throughout this paper.

Most of the observations were taken with a $\sim$1\,second integration time to track any small-scale phase fluctuations for longer baselines, although all band 3 data, some band 4 data (UID suffix Xa1e, Xc50, Xead, and X5d0 of SDP.81 EBs) and some band 6 data (UID suffix X1481, X11d6, X8be, and X1716 of SDP.81 EBs) use the standard $\sim$6\,second integration time. As the WVR data are also recorded on $\sim$1\,s timescales, WVR correction provides the opportunity to remove very short time variations, if they are present.

For all observations the bandpass source is observed for $\sim$5\,minutes, except in the Mira band 6 EBs, where it was observed for $\sim$10\,minutes. The WVRs were functioning throughout and the phase referencing scheme cycled between the target and phase calibrator more rapidly than in standard observations (order of minutes). Typically the on source (i.e. the science target) time for each scan ranges from $\sim$60 to 80\,seconds, while that spent on the phase calibrator is $\sim$15 to 18\,seconds, resulting in $\sim$75 to 100\,second cycle times. Only for the SDP.81 band 7, EBs were the phase calibrator scan times closer to 10\,seconds.

In order to quantify the effects of the WVR application and the effect of the scaling factor, we first calibrated the data following the reduction scripts supplied with the SV data via {\sc casa} \citep{McMullin2007} with the standard WVR application without a scaling factor applied. These standard calibrated datasets were flagged for errors as noted in the provided SV scripts. The optimal scaling factor for the WVR solutions was then found from the analysis of the phases extracted from the bandpass calibrator (see Sect. \ref{meth}). Subsequently the delivered reduction scripts were edited to implement the scaling in the {\sc wvrgcal} routine and then re-run. We imaged the bandpass calibrator (Section \ref{bpim}) from datasets where coherence improvements were indicated (39 of 75) and we also imaged the science target in cases where the scaling had a positive impact (Section \ref{imag}). Our target source imaging scripts essentially follow those provided with the SV delivery, although each EB is imaged separately and we did not apply time or frequency averaging (see Sect. \ref{allim}). Table A.1. in the Appendix lists the 75 EBs used for the analysis along with the weather parameters as extracted from the weather station metadata.

\section{Methodology and analysis}
\label{meth}

For a thorough investigation of the atmosphere, observing conditions, and to establish whether a scaling of the WVR solutions improves the data on various timescales, one would ideally require a long (tens of minutes) stare at a quasi-stellar-object (QSO) to establish the spatial structure function before and after WVR corrections \citep[e.g.][]{Matsushita2017} with and without WVR scaling. The information from such observations could be used to better correct science data, however, in reality a long stare at a QSO would introduce unacceptable overheads into observations.

However, for all science datasets the observation of the QSO at the start of the observations, targeted as the bandpass calibrator, can be used for a similar analysis. When using the bandpass calibrator data we must limit ourselves to examining only timescales up to $\sim$2-3\,min considering that the observations are only 5\,min long; i.e. we must sample this time range at least twice. We used the two-point-deviation (TPD) function to investigate the phase variations, $\phi_{\sigma}(T)$. This statistic in general allows one to investigate various timescales, ranging from the integration time to $t_{obs}/2$.  It also allows the isolation of certain timescales on which the largest phase fluctuations occur in comparison to a phase RMS measure, $\phi_{rms}$, which is simply an ensemble average of all the phase variations occurring for all timescales less than an adopted averaging timescale (usually the maximal observing time when used in atmospheric SSF studies). The TPD, at a given timescale, is the measure of phase variations or noise that we act to minimise by scaling the WVR solutions. The phase RMS is used later as a means to calculate coherence losses where both the entire observation time and a 60\,s averaging time are used (see Section \ref{phaseres}). 

\subsection{Two-point-deviation analysis statistic}

The two-point-deviation $\phi_{\sigma}(b,T)$ that we calculate is a function of baseline length, $b$, and time interval of interest, $T$, it can be defined by

\begin{multline}
  \label{eqn3}
    \phi_{\sigma}(b,T) = \bigg(\frac{1}{2(N-1)}\, \sum_{i=0}^{N-2} \\
     \times (\bar{\phi}(b,T,t_{i}+T)-\bar{\phi}(b,T,t_{i}))^2 \bigg)^{1/2},
 \end{multline}


\noindent where $\bar{\phi}(b,T,t)$ is a two element interferometric phase with baseline length $b$ averaged over the time interval $T$, starting at time $t_i$, and $\bar{\phi}(b,T,t_{i}+T)$ is an average of the phases (on the same baseline) also over a time $T$ but starting at time $t_{i}+T$. The value $N$ is the number of samples of duration $T$ in the phase stream \citep[e.g.][]{McKinnon1988}.\ The value $T$ is chosen to examine differing time intervals from the same phase stream by dividing into different subsets. In Equation \ref{eqn3}, one has $t_{obs}/T$ samples that are dependent on $T$, hence there is greater uncertainty associated with longer time intervals with an extreme case that $T = t_{obs}/2,$ which provides only $N$=2 samples. This is called the fixed time estimator, for example if $T$=6\,seconds then the first averaged phase, $\bar{\phi}(b,T,t_{i}+T)$, is an average of 6 seconds of data taken at $t$=6,7,8,9,10,11\,seconds, and $\bar{\phi}(b,T,t_{i})$ are the phases averaged at times $t$=0,1,2,3,4,5\,s (if the integration time is 1\,second); thus the first $N$ sample is the difference of these averaged phases. The next $N$ sample is provided by the difference between the phases averaged at times $t$=12,13,14,15,16,17\,s ($\bar{\phi}(b,T,t_{i}+T)$) and $t$=6,7,8,9,10,11\,s ($\bar{\phi}(b,T,t_{i})$), i.e. the jump between consecutive $N$ samples is $T$.

The phase stream data can be optimised and the noise reduced if we use an overlapping estimator, given by

\begin{multline}
  \label{eqn4}
  \phi_{\sigma}(b,T) = \bigg(    \frac{1}{2\,T(M-2T+1)} \\
  \times \sum_{i=0}^{M-2T}\bigg( \sum_{j=i}^{i+T-1}({\phi}(b,t_{j}+T)-{\phi}(b,t_{j}))^2\bigg)   \bigg)^{1/2}.
\end{multline}

\noindent Here $M$ is the number of phase elements ($\phi$) in time. Starting from $i$=0 to $M$ = $t_{obs} - 2T$ means the outer summation contains $M$ = $t_{obs} - 2T + 1$ samples of the inner loop (due to a zero indexing), which contains $T$ samples itself. For example, with $T$=6 as the time of interest in Equation \ref{eqn4} and for the first outer loop, $i$=0, we consider the phase differences between ${\phi}(b,t=6) - {\phi}(b,t=0)$, ${\phi}(b,t=7) - {\phi}(b,t=1)$, ${\phi}(b,t=8) - {\phi}(b,t=2)$, ${\phi}(b,t=9) - {\phi}(b,t=3)$, ${\phi}(b,t=10) - {\phi}(b,t=4),$ and ${\phi}(b,t=11) - {\phi}(b,t=5)$ (6 samples) as $j$ runs from 0 to 5. If we shift the phase stream up by one integration time (1\,sec), $i$=1, we now consider the phase difference ${\phi}(b,t=7) - {\phi}(b,t=1)$, ${\phi}(b,t=8) - {\phi}(b,t=2)$, ${\phi}(b,t=9) - {\phi}(b,t=3)$, ${\phi}(b,t=10) - {\phi}(b,t=4)$, ${\phi}(b,t=11) - {\phi}(b,t=5),$ and ${\phi}(b,t=12) - {\phi}(b,t=6)$. Thus in total we have $T(M-2T+1)$ samples ($>>$ N samples from Equation \ref{eqn3}). We explicitly note once here that this is the two-point-deviation, which is an Allan deviation without a weighting for the time interval $T$ (in an Allan deviation the divisor in Equation \ref{eqn4} would be $T^2(M-2T+1)$ and, for consistency, $T(N-1)$ in Equation \ref{eqn3}).

Both equations above provide the same results although the noise is greatly reduced for longer timescales using the latter. Equation \ref{eqn4} is therefore used throughout the analyses in this work.

\subsection{Data processing and diagnostic plots}
The data processing for extracting the phases, calculating the statistics, and testing the optimal scaling (described below) are fully automated in our {\sc python} package\footnote{http://www.alma-allegro.nl/wvr-and-phase-metrics/wvr-scaling/} created for the ALMA community. In this work for these SV data we follow a semi-automated approach to check each step individually before proceeding.

Most ALMA science observations will have at least one spectral window with a wide bandwidth to achieve good signal-to-noise on a phase calibrator and hence to allow phase referenced calibration. The maximal usable bandwidth in one spectral window generally ranges from $\sim$0.937\,GHz up to $\sim$1.875 GHz depending on the observation band and specific science spectral set-up. To mimic a typical science dataset taken in a mixed observing mode (TDM and FDM), our algorithm extracts the visibilities (i.e. phases) from only the averaged solution of the widest single spectral window from the dataset; in the case of TDM data the edge channels are already flagged.

\begin{figure*}
\begin{center}
\includegraphics[width=16cm]{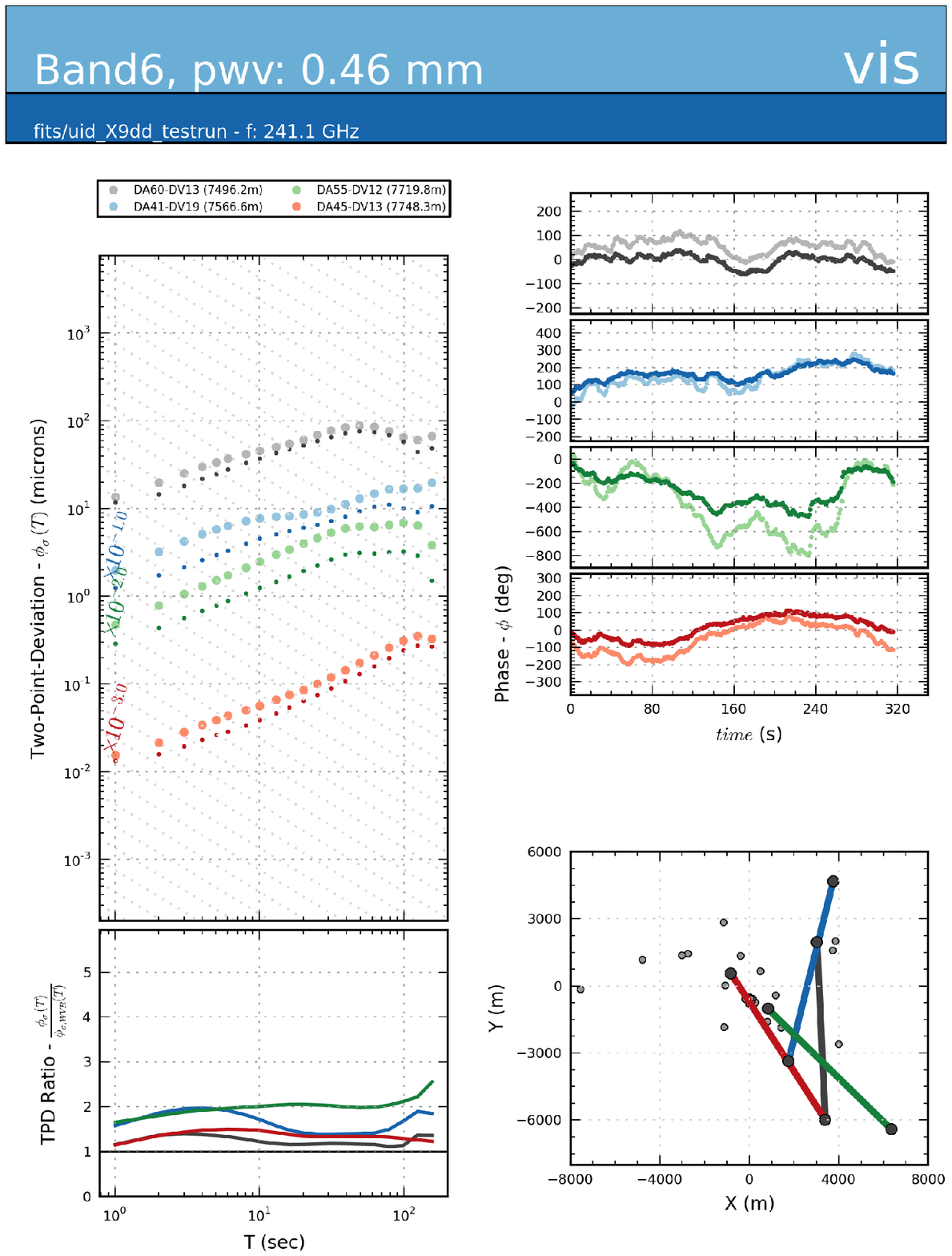}
\caption{One of the diagnostic output plots from the scripts reading in the datasets and unwrapping the phase streams. A key for the baselines and information about the dataset are shown
on the top left. The top left plot shows the two-point-deviations for the raw and corrected phases in micrometres (light and dark points, respectively; each baseline is scaled by an order of magnitude for clarity). The bottom left plot shows the ratio of improvement due to applying the WVR correction in a standard manner; for these band 7 data the average is $<$2. The top right plot shows the raw and WVR-corrected phase streams (light and dark) while the bottom right plot shows the positions of the antennas and associated baselines used. In the phase stream plots (top right) the scale of baseline DA55$-$DV12 is larger and the two-point deviations are also very elevated almost by an order of magnitude. This points to a possible issue with DA55.}
\label{fig1} 
\end{center}
\end{figure*}

Before starting the scaling analysis the phases are extracted from the visibilities and piped to an unwrapping algorithm. Unwrapping is required as interferometric phase visibilities are recorded between $-\pi$ and $\pi$ ($-$180$^{\circ}$ and 180$^{\circ}$), i.e. each antenna signal is within one wavelength of phase. Phase statistics cannot be calculated if the phase streams are not continuous in time. Although fully automated in our publicly available code, as noted above, we approach this as an iterative process as some phases were unwrapped incorrectly owing to either data or instrument problems given the nature of these SV datasets. Fig. \ref{fig1} indicates the type of plots generated; these plots allow the user to check the phase streams and two-point-deviation profiles for subsets of four baselines per figure. If any anomalous wrapping or baselines are found these can be noted to be corrected (if due to data errors that can be flagged) or can be ignored in the later scaling analysis. For the dataset shown in Figure \ref{fig1} there were no anomalous incorrect wraps, although a bad antenna was found where the baseline between antennas DA55 \& DV12 shows an anomalously large phase. These plots also show both the raw and standard WVR-corrected phase streams. A main point to emphasise is that there are residual phase variations, even on short timescales, for the WVR-corrected phases (darker symbols). The WVR correction is therefore not perfect. Also, the phase two-point-deviation statistics (left, Figure \ref{fig1}) are plotted in terms of path length noise to be frequency independent,

\begin{equation}
  \label{eqn5}
\Phi = \frac{\phi}{2\pi} \times \frac{c}{\nu_{obs}} (\mu m),
\end{equation}

\noindent where $\phi$ is the phase noise measured in radians, $c$ is the speed of light (in micrometres, $\mu$m), and $\nu$ is the observation frequency in Hz. For reference, a $\sim$30$^{\circ}$ phase RMS (corresponding to a 87\,\% coherence) corresponds to path length noise values of $\sim$250, 110, 70, 55, 38\,$\mu$m at 100, 230, 350, 450, and 650\,GHz.

\begin{figure}
\begin{center}
\resizebox{\hsize}{!}{\includegraphics{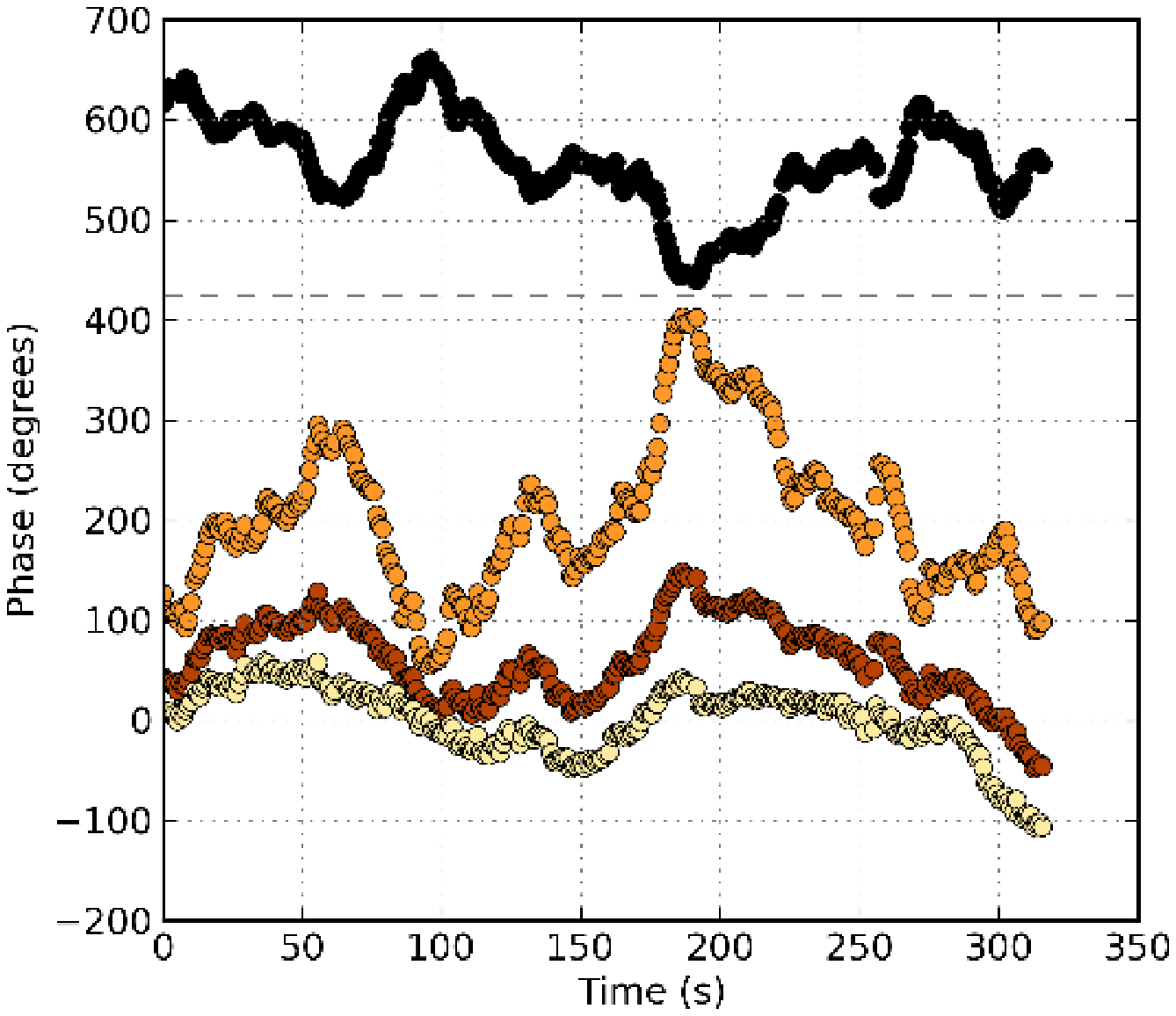}}  
\caption{Plots of the differential phase streams (degrees) for a $\sim$5500\,m baseline made between two antennas, DA51 and DV13, from EB X760. The orange symbols represent the raw phases as measured, the red symbols are the phases after the standard application of the WVR correction (unity scale), and the yellow symbols are the corrected phases after applying a scaling to the WVR solution, in this case the optimal factor 1.42 (Table B.1). The black symbols indicate the baseline-based WVR solution, which in the ideal case should exactly mirror the raw phase if only water vapour fluctuations (as measured by the WVRs) cause the phase delays in the raw phase. It is clear to see the amplitude of the phase fluctuations in the WVR solution (black) is less than those in the raw data (orange), thus leaving room for improvement via a WVR scaling factor.}
\label{fig2} 
\end{center}
\end{figure}

\subsection{Scaling the WVR solutions}

For the scaling analysis we use the raw unwrapped phases for each baseline extracted from the bandpass source and standard WVR solutions as created by {\sc wvrgcal}. A copy of the bandpass source visibility data are overwritten to that of a point source (a unity dataset where amplitude = unity and phase = zero). Subsequently the antenna-based standard WVR solutions are then applied (using {\sc casa}) to these unity data such that the correction applied data are the standard WVR solution$-$with a scaling factor of 1.0$-$but are essentially baseline-based. These solutions are scaled by a range of factors while they are applied to the initial raw phases (per baseline) in order to correct them. For each scaling factor, the TPD is calculated for the scaled, corrected data at a range of timescales ($T$ = 6, 12, 32, 64 seconds) shorter than the typical cycle time of these long-baseline SV data ($\sim$90 seconds) because referencing with the phase calibrator corrects the respectively longer fluctuations. Therefore, we only investigate the timescales that cannot be corrected with the phase referencing scheme. Our algorithm searches for the WVR scaling factor associated with the lowest TPD of the corrected phases, i.e. those with the lowest fluctuations, and further acts to re-adjust the WVR scaling factor (within a narrower range) until the raw phases have been corrected optimally and the TPD is fully minimised. The scale factor at each timescale that acted to minimise the TPD of the corrected phases is then reported and later the value is inputted manually in {\sc wvrgcal} during the re-reduction of the entire dataset where the WVR scaling applied. The range of the scaling values is capped between 0.05 and 2.5 with the smallest scaling increment of 0.01. The upper limit can be adjusted in our publicly available code although we do not find any cases where the scaling needs to be in excess of $\sim$2. 

In this work we examine the scaling for baselines made -only with the reference antenna- and -all the baselines in the array-. Owing to the analysis of significantly more baselines the latter is over 10 times slower (e.g. 5\,min versus 50\,min for the scaling analysis code). First however, the data must extracted from the raw delivered data and the intermediate files produced, this step itself can take on the order of tens of minutes to an hour for a typical dataset (with a 5- to 10-min-long bandpass) running on a standard desktop machine. No user input is required for the intermediate steps.

In Fig. \ref{fig2} we plot the raw data phase stream for X760 from a baseline between two antennas separated by $\sim$5500\,m (orange), the standard (scale = 1.0) WVR solution as extracted from the unity dataset (black), the standard WVR-corrected data (standard solution applied to the raw data; red), and the optimally corrected WVR scaled data (scaled WVR solution, factor 1.42, applied to the raw data; yellow). In an ideal case the raw phases from the astronomical source would only be corrupted by the wet (water vapour) content of the troposphere such that the WVR should be able to fully correct the corrupted data, i.e. the WVR solutions (black) should be an exact mirror of the raw phase stream itself. Although the WVR solutions are a close mirror to the raw data (Fig. \ref{fig2}, black = WVR solution and orange = raw phase) they are not exact, which leaves room to improve the WVR solutions. Notably, in Fig. \ref{fig2} it is clear that the amplitude of fluctuations in the standard WVR solution (black) are much lower then those in the raw phase data themselves (orange), and that the scaled WVR-corrected data (yellow) have a lower variability than the standard WVR-corrected data (red).

\begin{figure*}
\begin{center}
  \includegraphics[width=15.5cm]{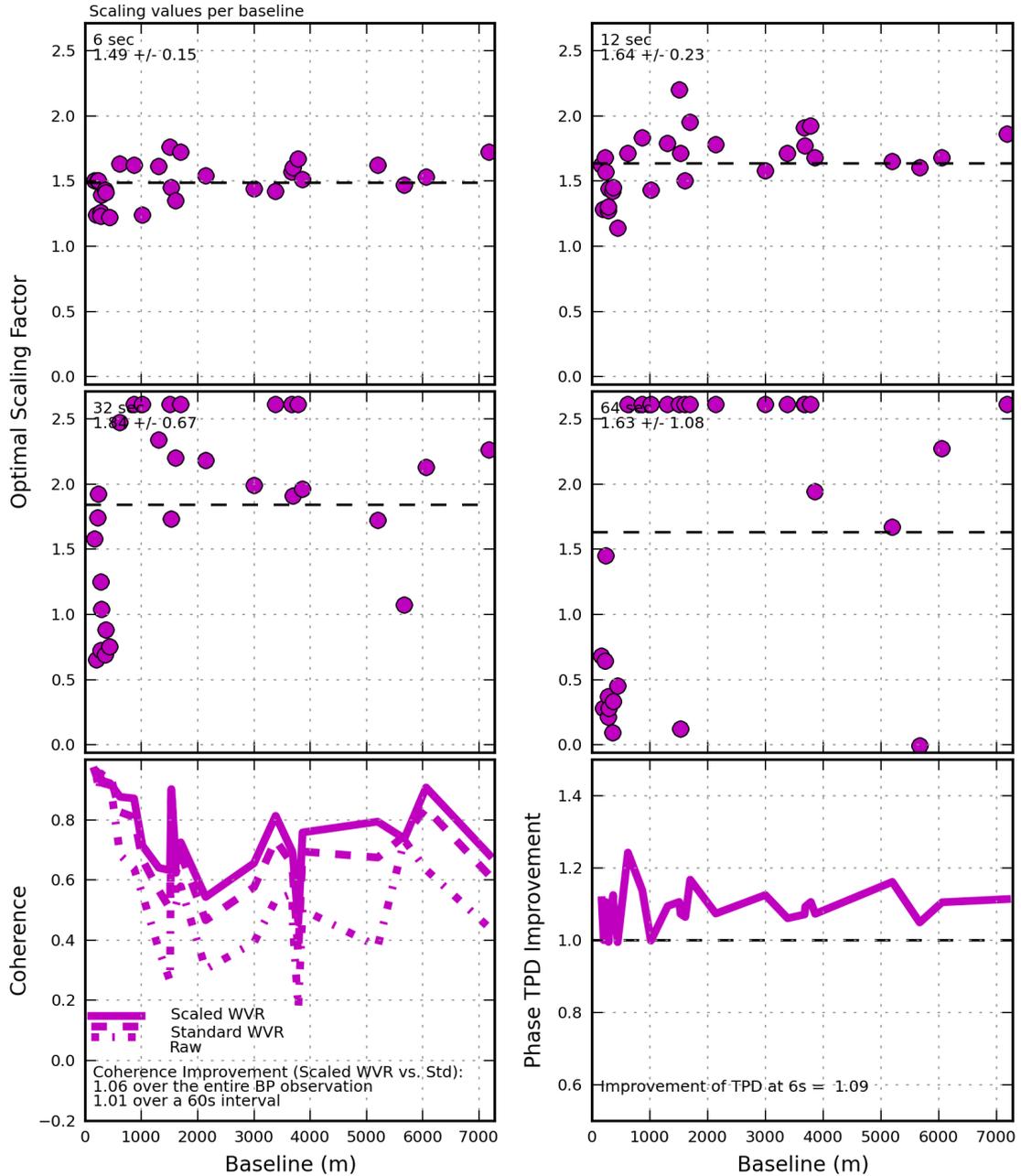}
\caption{Water vapour radiometer scaling diagnostic plot. The four plots, from left to right, top and middle rows are the optimal scaling factors vs. baseline length made with the reference antenna that minimises the corrected phase after the WVR application, according to the TPD phase statistics measured at 6, 12, 32, and 64\,second timescales. The bottom right plot indicates the additional improvement to the phase TPD after applying the average scaling factor derived for the 6 s two-point-deviation $\phi_{\sigma}(6s)$, while the left bottom plot indicates the coherence against baseline length for the raw, standard WVR-corrected and scaled WVR-corrected data. The coherence improvement expected over the entire observation of the bandpass (using the phase RMS from the full observation) and that expected over only 60\,s (related to the on-source time of the science target, using a phase RMS over a 60\,s interval) are reported. See http://www.alma-allegro.nl/wvr-and-phase-metrics/wvr-scaling for an example of all baselines plots.}
\label{fig3} 
\end{center}
\end{figure*}

Figure \ref{fig3} shows the WVR scaling factors versus baseline lengths that minimise the TPD, $\phi_{\sigma}(T)$ for $T$ = 6, 12, 32, and 64 seconds timescales (top$-$left to middle$-$right) in the corrected phase data (using baselines made only with the reference antenna). These plots are made for all EBs and are discussed in more detail in Sect. \ref{res}. This is the only diagnostic plot required to establish the best WVR scaling factor to apply to data. 

Our method of generating an intermediate baseline-based WVR solution for scaling is the most efficient and produces the exact same results as the alternatives. The other methods are using either many calls to {\sc wvrgcal} within {\sc casa} to create many WVR antenna-based solutions of various scaling factors or scaling the original antenna-based {\sc wvrgcal} solutions analytically and then applying them in {\sc casa}. Both of these alternatives require intermediate steps that must use multiple calls to {\sc casa} tasks to generate or apply the solutions and correctly interpolate them to the phase data for each different scaling factor trialled, before the `corrected' phases can even be extracted for the TPD analysis in order to run the minimisation. Typically these {\sc casa} tasks can take a few minutes each, which would snowball to hours in the course of the TPD analysis and minimisation where many tens to hundreds of scaling factors are tested. This makes any analysis using the {\sc casa} tasks prohibitively time expensive and hence our method only requires a single call to such tasks. 

\section{Results} 
\label{res}
This section details all results from the phase data analysis and the images made from each EB. Comparisons are also made of the established scaling factors with the data improvement values and with the weather and observational parameters. The improvement ratio measures used are detailed in the following subsections.

\subsection{Scaling factors}
\label{phaseres}

Table B.1. in the Appendix shows the results for the optimal scaling values averaged over -only the baselines made with the reference antenna- whereas Table B.2. shows the scaling values that were determined using the phases extracted from -all baselines in the array- between all antennas. The columns indicate the average scaling to minimise the $T$ = 6, 12, 32, and 64 second two-point-deviations (TPDs), $\phi_{\sigma}(T)$, the improvement ratios of the phase variations after applying either the 6 or 12\,s scaling factor (see below), and the coherence improvements expected. The latter coherence improvement measures are those that directly relate to any image improvement; the former ratio, i.e. the improvement of the phase noise, is not used in any comparisons as it does not map linearly with coherence improvements and hence cannot be used to predict the image improvement directly.

The coherence is calculated from Equation \ref{eqn0} for each baseline via the phase RMS. The improvement reported is the average value established using a ratio of the coherence calculated from the WVR scaled corrected data with the standard WVR-corrected data. Two estimated coherence improvements are reported: those related to the use of a phase RMS measured during a 60 second period of overlapping samples of the phases from the bandpass source (to relate to the expected improvement during the on-source time for the science target) and also those measured over the entire bandpass observation ($\sim$5\,min) to relate specifically to the improvement expected for the images of the bandpass (see Section \ref{bpim}).

\begin{figure*}
\begin{center}
\includegraphics[width=17cm]{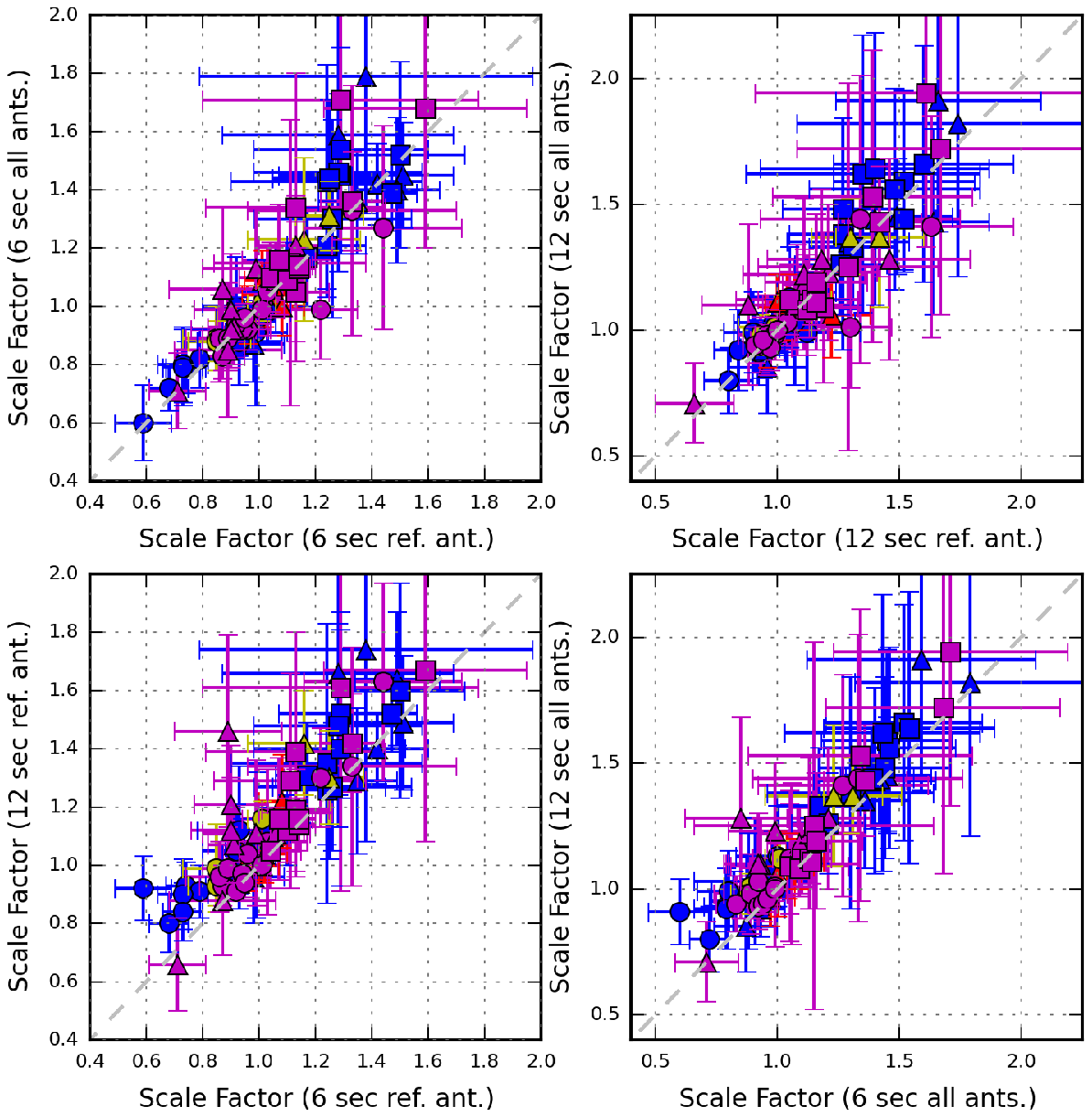}
\caption{Various comparisons of the scaling factors at the 6 and 12 second timescales as calculated for baselines made with the reference antenna only and those using all baselines in the array. The top panels show that for both the 6 (left) and 12 (right) second timescales the scaling factors calculated over baselines with the reference antenna and all antennas are coincident given the uncertainties in the scaling factors themselves and follow the 1:1 line (grey dashed). Comparing the 6 and 12 second scaling factors for baselines with the reference antenna (left) and those from all baselines (right) show that the 12 second factors are either $\sim$equal to those at 6 seconds or very slightly larger. Within uncertainties the scaling factors from the various calculations are typically in reasonable agreement. The colours represent the different sources, HL Tau in blue, Juno in red, Mira in yellow, SDP.81 in purple, while the symbols represent the different observing bands, band 3 or 4 are circles, band 6 are triangles and band 7 are squares.}
\label{fig4} 
\end{center}
\end{figure*}

The scaling factors we find vary from $\sim$0.6 to $\sim$2.0 for ($T$= 6 and 12\,s). Accounting for the uncertainties (represented by the standard-deviation), the scaling factors for all timescales per EB are reasonably consistent. They are also coincident between those found using baselines with the reference antenna only and those found while assessing all baselines when accounting for the uncertainties (Fig. \ref{fig4}). Furthermore EBs that indicate over 1\% coherence improvement in the reference antenna only analysis also indicate a similar improvement when using all baselines in the array. The latter analysis is more robust given the increased number of baselines that are analysed, however the trade-off in examining -only the baselines made with the reference antenna- compared to -baselines made between all the antennas- is that the former takes over a factor of 10 less time to run ($\sim$3-5\,minutes compared to $\sim$30-50\,minutes on a typical desktop machine for a single EB). Also the 12 second timescale factors appear to be slightly larger than those established at the 6 second timescale. We surmise that it could potentially be a real atmospheric effect in that the longer timescales trace larger fluctuations that require a slightly higher scaling to be optimally corrected. Furthermore, in some EBs the longer baselines sensitive to larger, longer timescale fluctuations also have 32 and 64 second scaling factors that are slightly larger. However, as the length of the observations are too short and the number of longer baselines is limited we cannot test such a hypothesis further with the available data.

The uncertainties are noticeably lower for the smallest time intervals, 6 and 12 seconds, which have more samples and the scaling factors per baseline are also more tightly constrained about the mean (also see Fig. \ref{fig3}). Either the 6 or 12 second scaling factor is selected for use in further analysis and imaging on the basis of selecting the factor that provides the most significant correction. In Tables B.1. and B.2., the EBs highlighted with `*' use the 12 second timescale scaling factor. In Table B.1. only, any EB with `**' are those where the reference antenna only scaling analysis did not find a positive improvement and the all baseline analysis factor was used instead (if an improvement was reported). The default was to use the 6 second reference antenna values, although as discussed most scaling factors are consistent at with 6 or 12\,s (see Figure \ref{fig4}).

\begin{figure*}[ht!]
\begin{center}
\includegraphics[width=17cm]{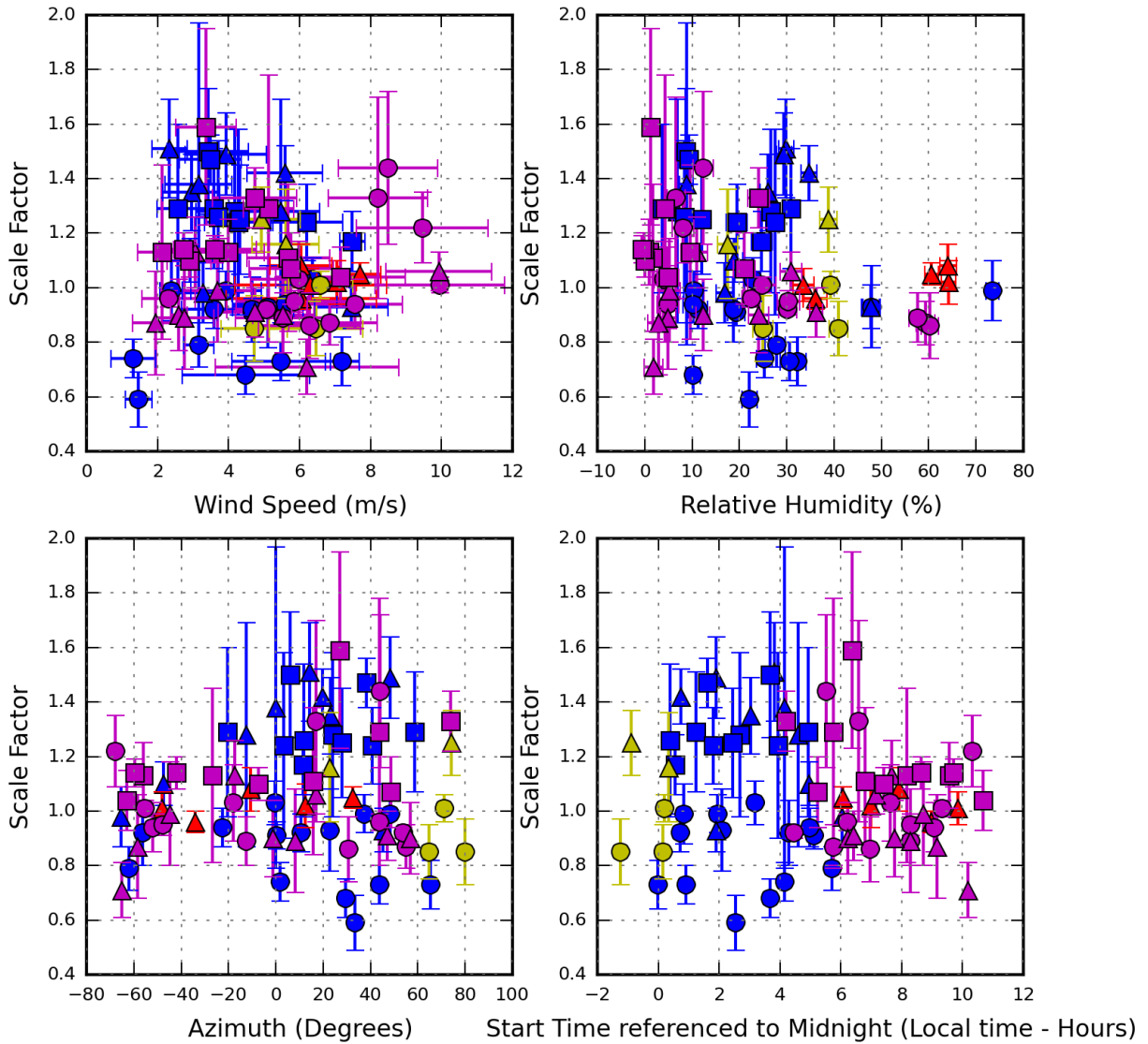}
\caption{Plots of the scaling factor as established from optimising the 6 second two-point-deviation statistic on baselines made with the reference antenna against wind speed, humidity, source azimuth, and start time referenced to midnight at the Chilean local time (CLT), where CLT = Coordinated Universal Time (UTC) minus 3\,hours. The colours and symbols are as Figure \ref{fig4}.}
\label{fig5} 
\end{center}
\end{figure*}

\begin{figure*}[ht!]
\begin{center}
\includegraphics[width=17cm]{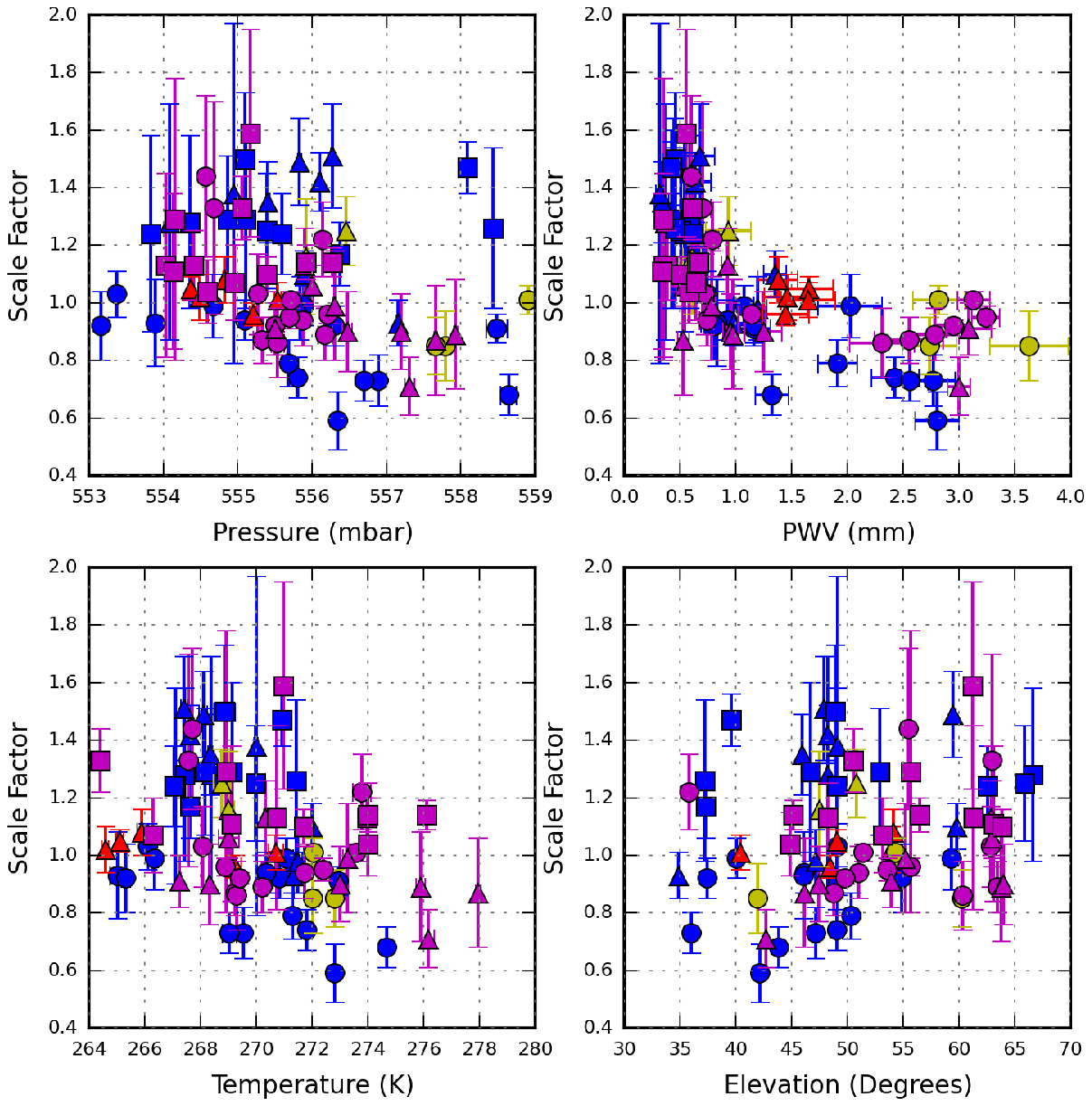}
\caption{Plots of the scaling factor as established from optimising the 6 second two-point-deviation statistic on baselines made with the reference antenna against pressure, PWV, temperature, and source elevation. The colours and symbols are as Fig. \ref{fig4}.}
\label{fig6} 
\end{center}
\end{figure*}

Our definitive result is that in most cases the factor for optimal phase correction is not unity and in a number of cases using the scaling factor in {\sc wvrgcal} produces better residual phases after scaled WVR correction. There are 62 EBs of the 75 that indicate the TPD statistic of the phase fluctuation is improved (i.e. improvement ratio $>$1.00); however, as noted above, this does not necessarily translate to a coherence improvement. A reduced amount, 39 of the total 75 EBs, have a coherence improvement as calculated from the phase RMS of the bandpass source over the entire observation time (21 of 75 if the improvement is $>$1.01), while only 12 EBs indicate there is an improvement on shorter timescales estimated by the coherence calculated from the phase RMS over 60\,s (i.e. $\sim$related to the time on the science target between phase calibrator visits, which calibrate out any longer term phase variations). Notably, only three of the low frequency EBs (bands 3 and 4) have any estimated coherence improvement (ratio $>$1.00), while the most prominent and numerous improvements appear to be associated with the band 6 and 7 observations (see Section \ref{conds2}).

\subsubsection{Water vapour radiometer scaling relationship with conditions}
\label{conds}
\begin{figure}
\begin{center}
\includegraphics[width=8cm]{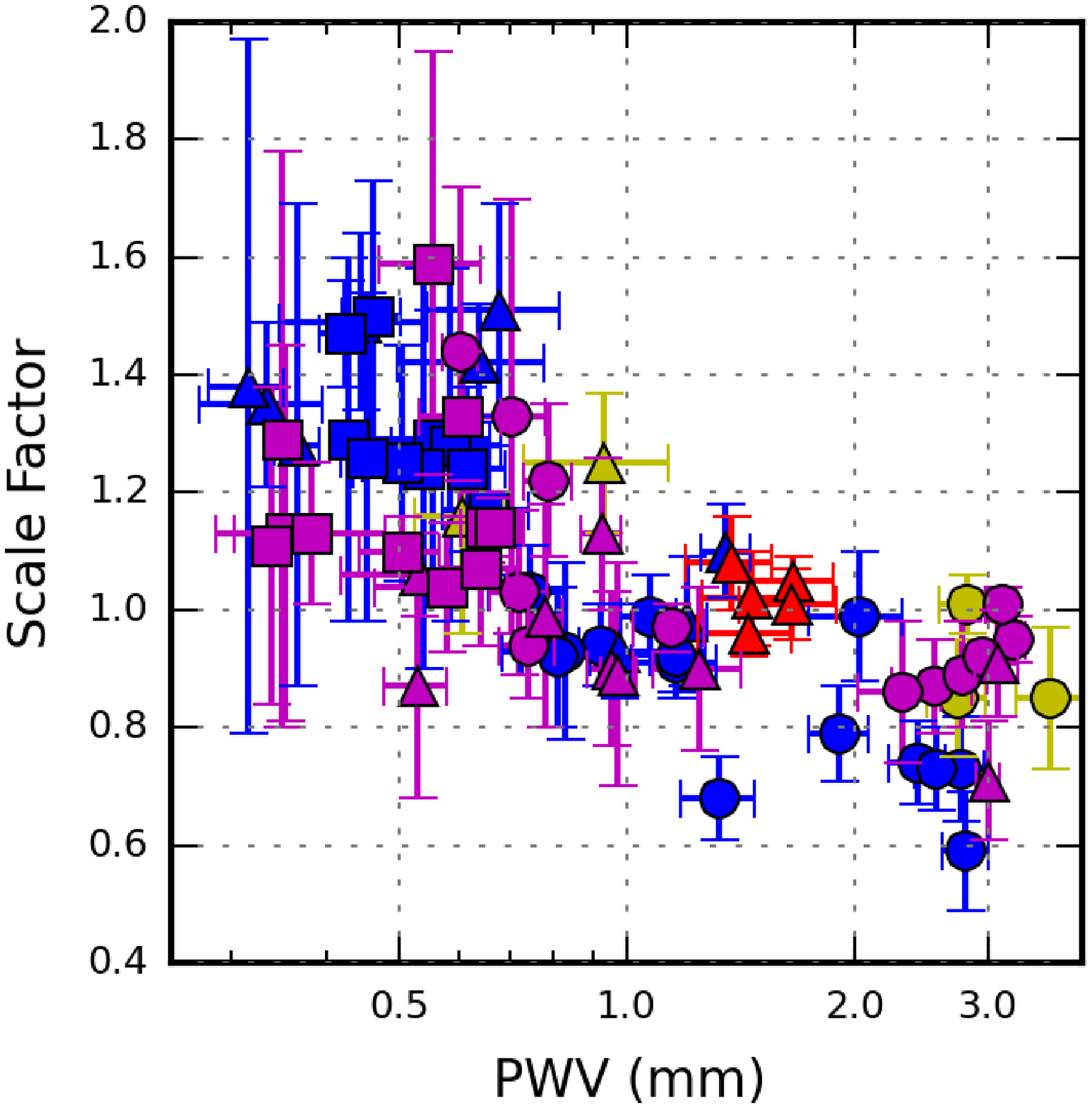}
\caption{Plots of the scaling factor as established from optimising the 6 second two-point-deviation statistic on baselines made with the reference antenna against PWV. The PWV is plotted on a logarithmic axis to highlight the trend as seen in Figure \ref{fig6}. The colours and symbols are as Fig. \ref{fig4}.}
\label{fig6addedLog} 
\end{center}
\end{figure}

We compare various weather and observational parameters with all 6 and 12 second timescale WVR scaling factors from both the reference antenna only and the all baseline analyses. Considering the weather condition parameters averaged over the observation time, the scaling factors do not have any correlations with the wind speed, humidity, source azimuth, source elevation, or observation start time when undertaking a Spearman rank test; Figs. \ref{fig5} and \ref{fig6} shows some of these quantities plotted with respect to the 6 second reference antenna scaling factors. There does appear to be some relationship with pressure, temperature, and PWV. The most visually clear is the relation to PWV; Fig. \ref{fig6addedLog} shows these on a logarithmic scale with respect to the 6 second reference antenna factors. The Spearman rank correlation coefficients ($\rho$) are $-$0.38, $-$0.39, and $-$0.76 for pressure, temperature, and PWV, respectively, when using the 6\,second reference antenna scaling factor. The significance of a given $\rho$ value depends on the sample size. For the 75 datasets a correlation $\rho$ of $\pm$0.355 is significant at the 99.9\,percent level (Table 3 of \citealt{Ramsey1989}), i.e. the probability of a null hypothesis is 0.1\,percent. Therefore we interpret the correlation with PWV as very strong, whereas those with pressure and temperature can be considered as `medium strength' correlations. Following equation 13.20 from \citet{Thompson2017} we compare the total, dry, and wet excess path lengths (Zenith and line-of-sight, accounting for elevation) with the WVR scaling factor. The excess path length calculation somewhat incorporates the measures of PWV, temperature, pressure, and relative humidity rather than using the single parameters alone, but still assumes a constant water vapour scale height and an isothermal atmosphere (see \citealt{Thompson2017}). We however do not find any significant correlation between the excess path length values and the WVR scaling factors.

Dividing the EBs into high and low PWV datasets, we see for EBs where PWV is $<$1\,mm that there is a much steeper relation between scaling factor and PWV (Fig. \ref{fig6} top right and Fig. \ref{fig6addedLog}). The Spearman rank correlation for this low PWV subset of 45 EBs ranges from $-$0.46 to $-$0.63 depending on which scaling factor value is used. This range is statistically significant at the 99.9\,percent level (the critical value is $\pm$0.45 for 45 datasets) and so we consider the correlation to be reasonably strong. The correlations between the other parameters noted previously are no longer significant. We emphasise that there are no significant correlations between any parameter and the scaling factor when using only higher PWV datasets ($>$1\,mm), even with PWV. Also, for the high PWV datasets, the scaling factor is generally less than 1.

Considering the variability of the wind speed, pressure, humidity, temperature, and PWV (as measured by the standard deviation) against the scaling factor we find that the larger scaling values ($>$1.2) only occur for the most stable conditions, where $\Delta$pressure $<$0.04\,mbar, $\Delta$PWV$<$0.15\,mm, and $\Delta$wind speed$<$1.5\,m\,s$^{-1}$. Although we find no significant correlations between the parameters, we offer two possible explanations for this phenomena: one suggests that in such stable and dry observing conditions there may be an underlying physical reason why higher scaling values are preferential, e.g. the dry and wet air fluctuations correlate in such conditions and therefore require a large scale factor to account for added delays (see Sect. \ref{disc_atmos}); the other, in contrast, is simply that for stable conditions the WVR corrections are so small (and have little effect) that one requires a larger factor to noticeably change the phases.

\subsubsection{Coherence improvement factor with conditions}
\label{conds2}

\begin{figure*}
\begin{center}
\includegraphics[width=19cm]{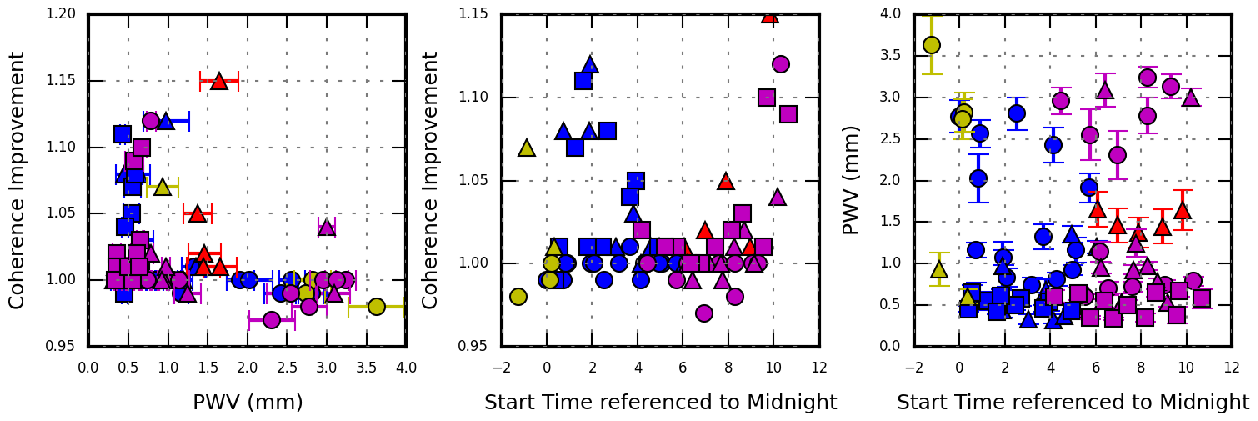}
\caption{Plots of the coherence improvement against PWV (left) and the observing time referenced to midnight (centre), and the PWV against the observing time (right). The coherence improvements are noticeably larger for lower PWV EBs, as well as possibly preferring earlier and later times ($\sim$midnight to 4\,am and after 9\,am, referenced to CLT, UTC - 3\,h). Any trend of improvement with time is not a result of only low PWV at these times, as there is no preference for lower PWV at these times; the symbols and colours are as Fig. \ref{fig4}.}
\label{fig6added} 
\end{center}
\end{figure*}

Figure \ref{fig6added} indicates the expected coherence improvement (considering the entire bandpass source observation) against the PWV and the time of day. We find that the majority of the improved EBs have PWV $<$1.5\,mm, although there is no correlation of PWV with coherence improvement directly. Generally higher PWV EBs appear to be worse overall in terms of coherence. There are also no other trends apparent between the coherence improvement and conditions except a minor separation with time. The EBs taken between $\sim$midnight and 4\,am and those taken after 9\,am appear to have elevated coherence improvement values that are larger than 3$-$4\%, whereas mid-morning EBs taken between 4\,am and 8\,am do not show such large improvements. This is not due to an underlying relationship of time with PWV, as there is a roughly homogeneous distribution of PWV compared with the observation start time (right - Figure \ref{fig6added}). Also, the low PWV datasets with $>$3$-$4\% coherence improvements are those with low wind speeds themselves ($<$4\,m\,s$^{-1}$). When comparing condition variability with coherence improvement we find that those data with larger improvements can be found in the stable PWV and wind speed conditions ($\Delta$PWV$<$0.15\,mm and $\Delta$wind speed$<$1.5\,m\,s$^{-1}$), however given the number of EBs with large $>$5\% improvement the results are not statistically significant.

Although a correlation of scaling factor with PWV exists, it does not translate directly to a correlation of coherence improvement with PWV and is not a linear relation. A simple parametric fit to estimate scaling factors cannot replace the analysis per dataset for each EB to find the optimal WVR scaling factor. Furthermore, for higher PWV data where the scaling factor is noted as less than one, although the phase noise is improved, the coherence overall is not. The investigation concerning relationships with observational conditions would clearly benefit from the analysis of many more datasets. In principle all parameters compared here can be extracted from any science dataset with at least 5 minutes time on a bandpass calibrator. Future investigations can therefore take place, but are beyond the scope of this paper focussing only on the long-baseline data. Moreover, the effectiveness of WVR scaling on other long-baseline data can only be tested with the long-baseline observations themselves, which began in November 2015 for Cycle 3 and will only begin to be publicly available in  2017.

\subsection{Image analysis}
\label{allim}
First we investigate the images of the bandpass calibrator to deduce whether the WVR scaling phase analysis and coherence improvement are as predicted. We then briefly discuss images of the phase calibrators followed by a more detailed discussion of the science target images.

\subsubsection{Bandpass calibrator}
\label{bpim}
Images of the bandpass source are made in all the cases where a coherence improvement was found (39 of the total of 75 datasets). Images were made with natural weighting with a shallow clean (50 iterations including a source mask of the central 15 pixels in radius) and also without cleaning at all (dirty images). A bandpass phase solution was applied after the WVR calibration (normal or scaled) but consisted only of a single solution value (per antenna) over the entire observing time of the bandpass source to correctly offset the average phase stream to zero degrees phase (interval = `inf' in {\sc casa}). We emphasise that the solution is -not- a self-calibration where the integration time would be used (interval = `int' in {\sc casa}). If self-calibrated phase corrections had been made after the WVR application, to correct the phases it would have invalidated any comparisons made to understand the impact of the WVR scaling. As the standard WVR correction is not perfect the phases are not exactly at zero phase, although they are distributed about zero phase once the single offset phase solution is applied and therefore the images are also imperfect. Any positive effects of WVR scaling should be reflected in the images of the bandpass source as the phase noise that caused any defects should have been reduced.

\begin{figure*}
\begin{center}
\includegraphics[width=19cm]{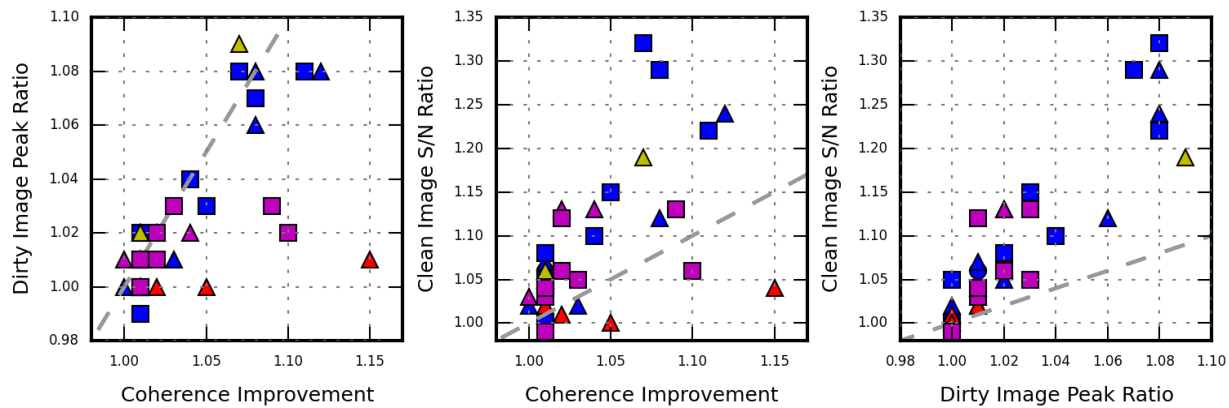}
\caption{Plot of the expected coherence improvement ratio as established using the phase RMS as measured over the entire bandpass observation compared to the ratio of the dirty image peak flux values of the bandpass source (left) and against the ratio of S/N values from the cleaned images of the bandpass source (centre). The right figure shows the ratio of the dirty image peak fluxes compared with the ratio of the S/N values from the cleaned images. The symbols and colours are as Fig. \ref{fig4} and the dashed line indicates a 1:1 relation.}
\label{fig7} 
\end{center}
\end{figure*}

Figure \ref{fig7} plots the expected coherence improvement with the ratio of the peak flux (WVR scaled applied / normal WVR applied) from the dirty images on the left and the ratio of the signal-to-noise (S/N) of the cleaned images (scaled / normal) centrally. The right panel compares the dirty image flux ratio with the clean image S/N ratios. The dirty images are used to compare the flux peaks as these are the `true' values unaffected by the deconvolution processes that occur during cleaning. In the cleaned images the S/N is used as this reflects both the increase in the peak flux value and also any decrease in image noise; reduced phase errors should better position the flux in the image whereas larger errors act to spread it around the image. The measured peaks, noise levels, and S/N values for each bandpass source imaged are listed in Table C.1. (see Appendix).

For the dirty images, we see there is a clear 1:1 correlation with the increase in peak flux compared with the expected coherence improvements, while for the cleaned images there is generally an elevated improvement that is above that expected from the coherence improvement estimate alone. As noted above, it is likely that the improved phases also help to position the flux in the image more optimally, thereby lowering the noise while increasing the peak flux. However an alternative or simultaneously effect could be occurring. Because we clean to a limited number of iterations, with a fixed gain in each step, the increased flux peak due to WVR scaling causes the cleaning process to clean more deeply per iteration and therefore reaches a lower noise value. We find that clean converges slightly quicker with WVR scaled data; fewer iterations, $<$50, for the WVR scaled images result in a noise level close to that in the standard WVR applied images with 50 clean iterations. Of the 39 datasets with an expected coherence improvement for the bandpass source there are 33 that have image improvements (25 of these $>$5\%). There are four EBs with worse cleaned images after the WVR scaling. A possible cause is an underlying bad antenna in the data that actually becomes worse after WVR scaling. We do flag additional antennas during the scaling analysis because of problems we find, although they are not flagged out in the delivered data reduction scripts, which we leave unchanged except for the including the scaling factor in {\sc wvrgcal}. 

\subsubsection{Phase calibrator}
\label{phaseim}
Ideally one would like to have a second quantitive check to evaluate the improvement the WVR scaling factor would provide for the science target using the observations of the phase calibrator given it is much closer on the sky. In the case of these SV datasets the phase calibrators are observed for at most $\sim$18\,seconds in time per scan before spending the next $\sim$60 to 80\,seconds on the science target during the phase referencing procedure. As such the calibration of the phase calibrator to offset the phases to zero per baseline using the `inf' interval timescale in {\sc casa} provides one solution for each $\sim$15 to 18\,second timescale scan. Therefore the phase calibrators already have excellent coherence, meaning a few percent improvement in the phase RMS does not result in a noticeable coherence improvement as there is little variability in phase over such a short time period to better correct with WVR scaling. Without observing the phase calibrator for a longer time (matching the on-source science target time) we cannot assess the direct effect scaling would have on the science target at a more co-spatial location. In some EBs here the bandpass and the phase calibrator are the same source, thus the expected improvement established on the bandpass should directly translate to the science target. We discuss how the source separation angle effects the improvements in Section \ref{discuss_sep}.

\subsubsection{Science targets}
\label{imag}

Using the datasets that showed positive results in the bandpass imaging steps (33 of 39) we image only the continuum emission from the science source for the individual EBs. These images are produced with exactly the same clean parameters as delivered in the SV imaging scripts for the respective sources, i.e. the same clean threshold, weighting scheme (briggs robust -- \citealt{Briggs1995}) and multi-scale clean parameters. In some cases we use a smaller number of clean iterations. This is because the supplied image scripts are intended for interactive cleaning requiring a user to stop the cleaning manually based on the image residuals and thus would generally not continue automatically for the given, large number of iterations compared to our automated cleaning procedure. We clean automatically to allow each image made with the normal WVR or scaled WVR calibrated data to be cleaned by the same number of iterations to provide the fairest comparison. We also increased the image size to better understand if there is an improvement in the image noise as some of the delivered scripts did not fully image out to the primary beam edge.

\begin{figure}
\begin{center}
\includegraphics[width=8cm]{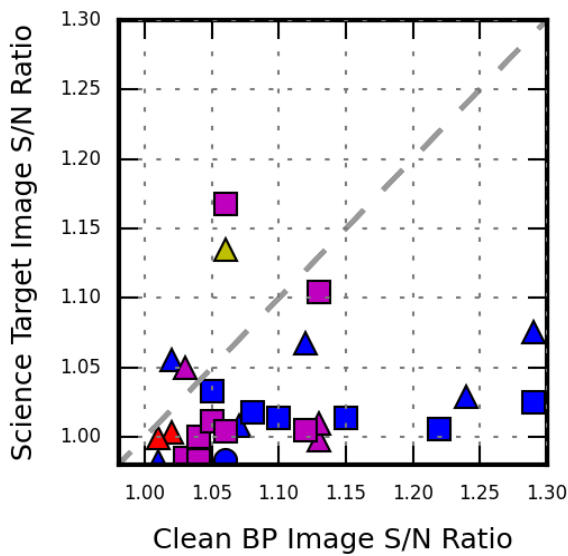}
\caption{Comparison of the cleaned bandpass source image improvement ratio compared with improvement of the science target images as measured by the ratio between the S/N of the images made with and without WVR scaled data. The Mira band 6 EB, X423, is excluded on the plot as it is improved by more that 50\,percent. The line indicates the 1:1 trend.}
\label{fig8} 
\end{center}
\end{figure}

Table D.1. lists the peak emission and the RMS noise as measured from the images where WVR scaling was and was not implemented. Although the bandpass images for X12c and Xa47, both SDP.81 band 4 data, showed some improvement, the science image from a single EB alone has a low S/N such that the source cannot be identified, and thus these data are excluded in Table D.1. From the 33 datasets, there are 23 improved, 6 have an improvement of less than 1 percent, 11 show between 1 and 5 percent improvement, while the remaining 6 show S/N improvements of greater than 5 percent. We cannot assess the two SDP.81 band 4 images. The magnitude of the improvement measured from the science images are in general larger than those expected from the coherence improvement calculated from the 60\,s phase RMS values of the bandpass phases (see Tables B.1. and B.2.) - at most 2 percent improvements were estimated.

Figure \ref{fig8} shows the comparison of the science target and bandpass source image improvement ratios measured by the ratio between the S/N of the images made with and without WVR scaled data. Comparing with the bandpass images the improvements are lower for the science targets. This however is not unexpected given the science data are corrected with phase referencing down to $\sim$60-80\,s timescales, whereas the bandpass has only one solution applied over 5\,minutes. Furthermore, the bandpass calibrators used to establish the scaling factors are not co-located on the sky and can be up to $\sim$27\,degrees away for these SV datasets. The scaling factor therefore may also vary in different lines of sight through the troposphere (see Sect. \ref{discuss_sep}). There are two images created per EB (as per the delivered SV scripts) for
Juno to track the rotation of the asteroid; in Figure \ref{fig8} the average improvement is plotted but both images are listed in Table D.1.

\begin{figure*}
\begin{center}
\includegraphics[width=17cm]{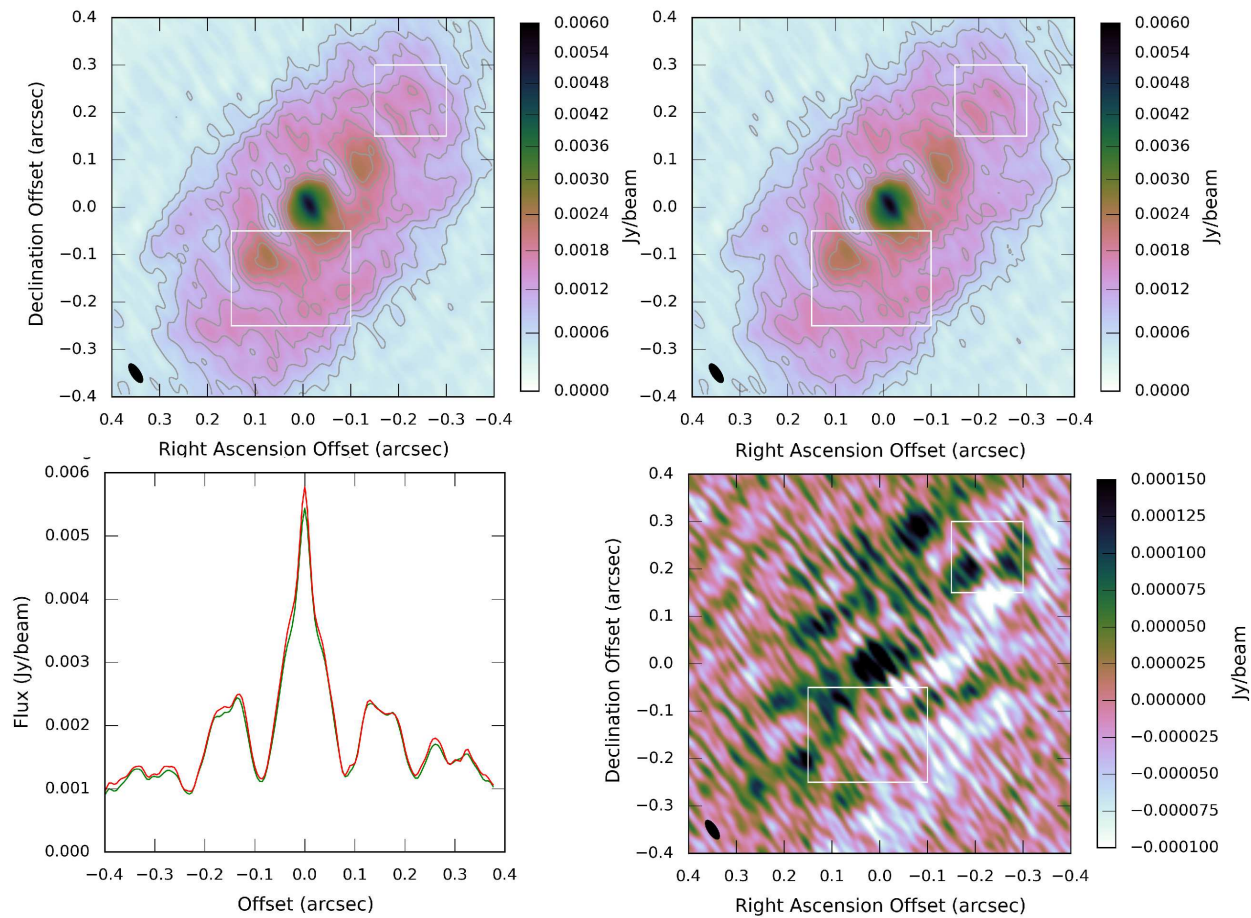}
\caption{Comparisons of HL Tau images from the EB X760. The top left and top right represent the images made with the standard WVR and scaled WVR solutions applied. The colour scales for both are fixed to a peak value of 6.0 mJy/bm, where the peak fluxes are 5.5 and 5.9 mJy/beam for the images with the standard and scaled WVR applications. The bottom right indicates the difference image (scaled $-$ standard), while a profile is shown in the bottom left (red indicates the WVR scaled image). The profile is extracted along a line from $\sim$(0.3,$-$0.3) to $\sim$($-$0.3,0.3). The boxes on the plots highlight regions to the NW and SE as discussed in the text.}
\label{fig9} 
\end{center}
\end{figure*}

The improvement of the Mira band 6 images, where both the peak flux increases and the map noise significantly decreases, is particularly worth mentioning. As the continuum emission structure of the Mira binary system is very simple \citep{Vlemmings2015} we do not see a noticeable change in terms of image fidelity. However, for some HL Tau images, specifically the X760 EB we find some very subtle changes in the extended continuum emission. Fig. \ref{fig9} shows the standard WVR-corrected image for HL Tau from EB X760 on the top left. The ring and gap-like structure (the gaps are not devoid of emission; c.f. \citealt{ALMA2015b}) are clearly visible and so the data can be deemed to be already reasonably well calibrated with the standard WVR application. Compared with the scaled WVR-corrected image (top right plot) the peak flux is found to be higher and the noise slightly reduced. Not only is the peak flux increased but additionally the contrast increases in the bright rings and gaps as can be seen in the profile cut in the bottom left panel of Fig. \ref{fig9}. Some other improvements are also highlighted by the two boxes. The ring-gap-ring emission to the north-west (also see the profile between 0.2 and 0.4 arcsecond) and  south-east of the peak is sharper. The difference image (bottom right) indicates that for the latter boxed region there is also a shift of the emission to a more central north-east position.

Positive image changes can potentially have an impact on the underlying science in other ALMA projects where WVR scaling helps to improve the S/N and contrast, especially if many improved EBs are combined in a final image. Hypothetically, considering other datasets that may not have phase referencing timescales as short as these SV data, an image improvement due to WVR scaling could easily mean the difference between a detection or not (e.g. considering the $>$10\% improvements seen over the 5\,min on-source time for the bandpass source; also see Section \ref{discuss_imps}). Overall however the general improvements in terms of fidelity of these science images are not significant enough to illustrate these possibilities. 

\section{Discussion}
\label{discuss}

It is clear that only a subset of the data are improved by WVR scaling. Investigating both the scaling factors and coherence improvement ratios for potential correlations with weather parameters indicates that low PWV $<$1\,mm and stable condition datasets can be improved. There is a known diurnal cycle at Chanjantor in terms of wind speed and temperature, for example \citep{Evans2003,Stirling2006}. Although the EBs as part of these data are taken on different days, we consistently see that data taken slightly before midnight (referenced to Chilean local time, UTC - 3 hours) to around 4\,am and those after $\sim$9\,am are those that can be improved the most and thus could point to a particular time where we could expect WVR scaling to improve data quality. The data taken at times responding the most to WVR scaling may point to a physical phenomena that is occurring.

Below we discuss potential causes of the scaling factor, the separation angle of the bandpass source and science target, and where the WVR scaling is most applicable; we also discuss the uniqueness of these SV data.

\subsection{Potential cause of the scaling factor}
There are numerous possibilities why a scaling factor is required and why the WVR alone does not make the most optimal solution in the first instance. These can be divided into instrumental or software problems and those that are atmosphere-based.

\subsubsection{Instrumental and software}
In these EBs the scaling (on short times 6 and 12\,s) is effectively constant with baseline length, suggesting that baseline-based instrumental effects, such as the correlation or the line length correctors would not be to blame. Antenna-based instruments could be a cause, such as receiver noise, related electronics, or indeed the noise in the WVR calibration with the hot and cold temperature loads (the hot load is actually `warm' at 80$^{\circ}$C). Slight variations in specification could cause extra phase noise. However any instrumental noise is not expected to be coherent with the real phase or the differential phase between antennas. One would not expect all instruments on all antennas to have the same noise issues, such that they would cause a systematic offset that can be alleviated with a scaling factor as we see here. The scaling values found in this work can be vastly different per day, which would not be the case if the same instrumental problems are causing the extra phase noise. If there were any instrumental problems based on temperature variability, these could show differences per day. However, the temperature change at each antenna, for each instrument, would have to be the same to cause the almost constant scaling factors seen per baseline, and therefore  this does not appear to be feasible.

Considering possible software assumptions, the correlation of PWV with scaling factor when we consider all EBs could point to a small error in the solutions calculated in the {\sc wvrgcal} code as a function of PWV. Phase variability is known to increase linearly with $\Delta$PWV, which is generally larger (but not always) when the PWV is greater itself \citep{Matsushita2017}, although it is difficult to isolate only PWV from the other variable observational parameters such as wind speed. If WVR scaling is the correction due to an error or assumption in the code then one might expect a clear linear or power-law relationship where the scaling factor was systematically dependent on PWV. Although the low $<$1\,mm PWV data follow a trend with scaling factor there is still considerable scatter that a systematic, computational error would not produce.

One possible issue does arise based on assumptions used in the software to model the atmospheric emission that is matched to the WVR signal. This model is `relatively simple' \citep{Nikolic2013} and consists of a single atmospheric layer and thus does not consider a thick water vapour layer. We caution that if the thickness of the layer is discrepant with the assumed model, it may relate to the absolute PWV and therefore would cause a scaling factor change with PWV as we find. Additionally, any scatter could be explained if the thickness of the atmospheric layer varies within a given range (for a given PWV) and causes variable deviations of the model with the measurements. Testing such a scenario to assess the effect of the WVR scaling factor is beyond the scope of this work  but should be considered as important future work.

\subsubsection{Atmospheric}
\label{disc_atmos}
Alternatively, atmospheric effects could be the cause of the scaling factor. Liquid water, in the form of fog or clouds, is known to be an absorber of the continuum electromagnetic radiation \citep{Ray1972, Liebe1989, Liebe1991}. Because the WVR use filter bands (or bandpasses; see \citealt{Nikolic2013}) the liquid free line model used for the atmosphere to generate the WVR solution would not work in these conditions \citep{Matsushita2000,Matsushita2003}. If liquid water was in the atmosphere in the form of clouds, then these may be more local given the long-baseline nature of the array configuration in these observations and may potentially affect only a group of antennas, specifically changing the scaling for only that group. This is not seen. Moreover, if there were a diffuse cloud over the entire array, one would not expect its internal structure to be homogeneous or correlated with the WVR measurable water vapour for each antenna such that the scaling would remain constant for all baselines. Clouds are mostly expected in high humidity and high PWV classically bad weather conditions, rather than in the conditions where the PWV is $<$1\,mm as in the EBs most improved in this work. There is also now an extra algorithm built in to the latest {\sc casa} code for use in conjunction with {\sc wvrgcal} to account for clouds in bad conditions ({\sc remcloud}), which acts to measure the underlying water continuum emission from the clouds. Here the cloud is removed under the assumption that it contributes only to the continuum emission in the WVR filters; it is not related to the shape of the water vapour line and only occurs in data where PWV $>$1.5\,mm.

Water ice could also be a cause of extra delays \citep{Hufford1991} although it is unclear if this is present in the atmosphere above ALMA. If ice were on the surface of the antennas the path length (delay) to each antenna could also change. However, for the amplitude of the differential phases to increase, as we account for by the scaling factor, all antennas in a baseline with the reference antenna would need the same change in ice layer thickness compared to the reference antenna. This is highly unlikely.

Dry air fluctuations in the troposphere could also be responsible. These are known to cause refraction but to a lesser extent than the water vapour in the millimetre/submillimetre regimes. In lower PWV conditions (e.g. $<$1.0\,mm PWV), under the assumption that the WVRs measure and correct all the water vapour caused delays, we can begin to contemplate whether the cause of the remaining phase fluctuations is the variability of the undetected dry air. If, for example, some of the dry air is correlated with the motions of the water vapour then the dry air would cause an additional refraction that would be the same as that induced by the water vapour, although it remains unmeasured directly. Mixing could in principle occur in all conditions: both when PWV is high and low, but the dry variability would only become more apparent in the case of low PWV ($<$1\,mm). The water vapour as measured by the WVRs could therefore be scaled to correct for the extra delays owing to the mixed-in dry air.

\citet{Stirling2005} performed simulations to investigate the dry and wet (water-vapour) fluctuations at Chajnantor. Specifically, the ALMA memo details the study of daytime fluctuations where the solar heating acts to create convective turbulent columns in the troposphere, i.e. temperature fluctuations cause the variability in the motion of the air. In the simulated conditions if the air close to the ground contains water vapour at early times then the dry and wet air fluctuation can become correlated. Later during the day there may in fact be an anti-correlation due to differences in temperature and moisture fluctuations such that the dry delays actually cancel with some of the delays caused by the water vapour. The authors note that at night-time, coincident with some EBs we noted to have the best reported image improvements, the potential for mixing between wet and dry components could be due to wind shear, however this simulation work was not completed. 

Using radiosonde data, \citet{Stirling2006} show the diurnal heating and cooling and varying wind speed at Chajnantor. They also established the temperature and moisture profiles as a function of height above the array. During the daytime the authors noted a strong gradient close to the ground due to surface heating as also used in their modelling \citep{Stirling2006}. These temperature gradients and fluctuations are those responsible for the turbulent air that leads to the mixing of the dry and wet air components. At night the profiles appear more complex but there are still large temperature gradients close to the ground and more importantly at the inversion layer at $\sim$1000\,m above ground. This is where the water vapour layer traditionally causing much of the phase instabilities is located and where temperature gradients could also act to mix the dry and wet components such that the total delays (dry and wet) are correlated with those of the wet, as measured by the WVRs.

We suggest that a mixing and correlation of the dry and wet air components provides a plausible explanation for the scaling factors found, especially when considering the data from $\sim$midnight to $\sim$4\,am (wind shear) and after $\sim$9\,am (ground-heating) show the best improvements. \citet{Stirling2005} show that in dry conditions where the wet path length drops $<$180\,$\mu$m, during $\sim$25\% at ALMA based on their radiosonde work, the wet and total fluctuations are correlated, on average by 0.75, leaving the remaining fluctuations to be caused by the dry component. Thus, making a very crude association we find that the WVR scaling for low PWV $<$1\,mm conditions, $\sim$1.2-1.3, applied to the wet fluctuations, can make up the `remaining' fluctuations in the driest conditions in the case where we assume that the wet and dry are indeed correlated. As we do not see any trends between variability of temperature with the scaling factor or the final improvements we slightly prefer the night-time wind-shear hypothesis for mixing. However, in these SV data we only have access to parameters from a single ground-based weather station data and therefore cannot examine any information on the temperature variability over or above the array itself. Future data with more weather stations active should be investigated to examine relationships to weather conditions and future work will be undertaken to model the wet and dry fluctuations over ALMA, during day and night cycles, to understand whether the scaling is caused by correlated wet and dry air.

\subsection{Source separation}
\label{discuss_sep}
In phase referencing a smaller phase calibrator and science target separation can provide the best correction of the phases through phase transfer \citep{Asaki2016}, thus we have to consider the variability of the WVR scaling factor with sky position. During any observation it is possible that different regions on the sky may be characterised by different scaling factors, as the antennas line of sight are through a different region of the troposphere, which in turn could behave differently. The examined EBs have relatively nearby bandpass calibrators typically within 10 degrees, but in a few cases out to a maximum separation of 27 degrees; see Table D.1. (PI science observations can use bandpass calibrators out to $\sim$120$^{\circ}$). Comparing all science images improvement ratios we find no relationships or correlations with the separation angle to the bandpass source used to establish the WVR scaling factor. Setting aside the small number statistics we consider that turbulent structures in the atmosphere are generally fairly similar, in terms of responding to WVR scaling, within a 30$^{\circ}$ region (maximum separation for these EBs). This does not mean the absolute phases are the same, only that the turbulent structures within the observing area are self-similar. However, for a more robust conclusion further datasets would need to be examined that have the addition of a longer ($\sim$5\,min) observation on the closer phase calibrator.

\subsection{Applicability of the scaling factor}
\label{discuss_imps}
The analysis of all datasets show that even in some cases with significantly different scaling values from the standard value, 1, and where the phases statistics are improved, the images themselves are not improved. The majority of these cases are the lower frequency EBs where such an outcome is somewhat expected as the coherence of the phase streams are already over $\sim$90-95\% because the phase RMS, in degrees, is noticeably lower at lower frequencies (and $<$30$^\circ$ in these SV data). Thus, even though the WVR scaling reduces the phase RMS, the coherence improvement is negligible. For higher frequency observations the weather conditions of these data improve, although the phase RMS in degrees also increases such that the coherence for higher frequencies is reduced overall. For higher frequencies, band 6 and to a further degree band 7 data, where WVR scaling improves the phase statistics, there is also generally an improvement in the coherence and therefore more noticeable image improvements. These improvements are also more apparent when using longer time intervals to measure the phase RMS over, i.e. comparing the 60\,s and full bandpass observation ($\sim$5\,min) coherence improvements, as the longer timescales also see larger phase fluctuations, which the WVR can correct. Because the science targets in these SV data are calibrated, by the phase calibrator, on 60$-$80\,second timescales compared to the assessment made on the bandpass source over 5\,minutes the images of the scientific target of interest are not improved as noticeably as those of the bandpass; the phase referencing corrects the longer timescale fluctuations.

Moving to even higher frequencies (band 8 $-$ 450\,GHz and above) where the phase variations, in degrees, increase further, the WVR scaling could have an even more noticeable effect. As the phase stability is both frequency and baseline dependent, classically shorter baseline ($<$5\,km) higher frequency observations ($>$band 8) have the same phase variations as the longest baseline band 7 data examined here that were shown to respond well to the WVR scaling. Currently the longest baseline data available (ALMA archive) for these frequencies are $\sim$2\,km and have $>$60\,s cycle times. Typically these data spend up to 5\,min on the science target and, as such, any phase variability occurring on shorter timescales have no means of correction other than the WVR scaling, excluding self-calibration using the integration time. Even in cases where self-calibration is possible for the targets, this may not always be possible for the integration time, thus only WVR scaling can correct the very short-term phases. For the longer cycle times the WVR scaling would provide a larger correction that is more consistent with what we find for the 5-min-long bandpass observations ($>$1-5\,\%). When moving to longer baselines for such frequencies, faster cycling to counteract the more rapidly changing atmosphere are required \citep{Asaki2016}. However, in these higher bands it is more difficult to find phase calibrators that are strong enough (S/N $>$15 in times $<$60\,s) and therefore  `fast' switching of the order 60\,s or less cannot be undertaken as 60\,s or longer observations of the phase calibrator itself are required to obtain a enough signal-to-noise for normal phase referencing. Again, longer cycle times only correct longer term atmospheric variations, meaning the WVR scaling would provide a much more noticeable improvement to the phases of the science target that may otherwise become decoherent.

\subsection{Uniqueness of the SV datasets}
\label{discuss_un}
Data used for this investigation are those taken with the intention to showcase the capabilities of ALMA. As such, interesting science targets were chosen that had reasonably close and strong phase calibrators. Moreover, reasonably fast switching was used (Section \ref{obs}) and the higher frequencies (bands 6 and 7) were reserved for the very best weather possible (both low PWV and where stability was excellent). For PI science observations, which have been and will be taken using the ALMA long baselines, such conditions in terms of weather and calibrators will not always be as ideal as those for the LBC SV data (Catherine Vlahakis, private communication) and thus the effect of WVR scaling could have a greater potential. From the discussion above we expect that in situations with weaker phase calibrators and increasing cycle times there would be a  larger margin for improvements from the WVR scaling (e.g. more comparable to the noticeably larger improvement of the bandpass images compared to the science target; 5\,min vs. 60\,sec timescales). Additionally, phase calibrators more distant on the sky ($>$3$^{\circ}$) have a less optimal phase transfer and thus scaling the WVR observed through the same line of sight as the science target may also provide a greater improvement. A future investigation of the PI science long-baseline data will be undertaken when the data is released (data taken late in 2015 should be available in 2017), these are also expected to have a wider range of conditions generally unbiased to the best cases used here during the SV testing.

\section{Summary}
\label{summa}
An investigation to optimise the WVR phase solutions has taken place using 75 ALMA long-baseline science verification datasets (EBs). Using the bandpass calibrators for each of the execution blocks it was shown that the standard application of the WVR corrections in {\sc wvrgcal} does not always result in the optimal correction of the raw phases. From the bandpass phase noise statistics, using a two-point-deviation calculation, an optimal scaling factor can be found, which when applied is shown to improve the phase correction of the entire dataset. Of the 75 datasets, 62 show an improvement in terms of phase noise statistics after WVR scaling, which translates to an improved coherence for 39 EBs as a result of the lower phase noise. Of the 39 EBs, 33 indicate an improvement in the image S/N, comparing the WVR scaled and non-scaled images of the bandpass calibrator. Of the 33 datasets with improved bandpass images, a reduced number of 23 show science target image improvements, in terms of S/N.

The application of the WVR scaled solution on the target generally results in a better image S/N and in a handful of cases can improve the contrast between features. Any changes in image fidelity are relatively small to the point of being unquantifiable. For the majority of the EBs analysed in this work the S/N typically increases by around 1$-$2 percent, but there were situations where significant increases, over 5 percent, are seen. These gains are slightly more than those expected from the estimated coherence improvement found using the phase RMS calculated from the phases of the bandpass calibrator over a 60 second timescale. We find that of the higher frequency datasets investigated, bands 6 and 7 respond to the WVR scaling, whereas lower frequencies (bands 3 and 4), which already have good phase coherence, do not. The scaling factor should ideally be deduced from a strong target as close as possible on the sky as the science source, although the bandpass source can be used if within 30\,degrees on the sky; any improvements could be limited, however.

Correlations with PWV were found when considering all the EBs together. We also found evidence to suggest that data taken around midnight to $\sim$4\,am and after $\sim$9\,am can be better corrected. The trends with PWV indicated a larger scaling value for lower PWV data and that in the dryer conditions, where PWV $<$1\,mm, scaling can have a more significant effect in improving image coherence. One promising hypothesis to explain the WVR scaling factor is that in certain cases there maybe mixing between the dry air constituents and the water vapour (wet) as measured by the WVRs. In this case if the dry air follows the same motions as the detected water vapour it would cause additional delays of the same `pattern' that can be accounted for by scaling the WVR solutions. Such mixing could be a result of temperature gradients or due to wind shear and therefore may occur preferentially at a certain time of night or early morning, as we find here. The examination of more datasets and detailed modelling is required to test such a hypothesis. Also, without more detailed investigations of the software underpinning {\sc wvrgcal,} we cannot rule out the assumptions within the atmospheric model used to find the WVR solution as the cause for the WVR scaling.

We also discuss the implication of cycle times and WVR scaling possibilities for higher frequency observations ($>$450\,GHz). Primarily, longer cycle times (at any frequency) cannot correct the short-term phase variations and thus leave a larger margin for the WVR scaling to improve the science target phases. Specifically, for higher frequencies fast ($<$60\,s) cycling may not be possible due to weak calibrators, which need more on-source time, and as a result the WVR scaling could offer a considerable improvement, i.e. $>$5-10\,\% to science images, as this provides the only means to calibrate the science target phases if integration time self-calibration is not possible.

\begin{acknowledgements}
LTM, RPJT, MRH, YC, and CT are part of Allegro, the European ALMA Regional Centre node in the Netherlands (also formerly TvK and MS). Allegro is funded by the Netherlands
Organisation for Scientific Research (NWO). This paper makes use of the following ALMA data: ADS/JAO.ALMA\#2011.0.00013.SV, 
ADS/JAO.ALMA\#2011.0.00014.SV, ADS/JAO.ALMA\#2011.0.00015.SV, and ADS/JAO.ALMA\#2011.0.00016.SV. ALMA is a partnership of ESO (representing its member states), NSF (USA) and NINS (Japan), together with NRC (Canada) and NSC and ASIAA (Taiwan), and KASI (Republic of Korea), 
in cooperation with the Republic of Chile. The Joint ALMA Observatory is 
operated by ESO, AUI/NRAO, and NAOJ. LTM and Allegro would like to thank
Catherine Vlahakis and Satoko Takahashi for their support during site visits, all parties involved in the ALMA/Allegro
Phase Metrics Meeting (held in Leiden May/June 2016), and other Allegri, Ian Stewart, Daniel Harsono and Ciriaco Goddi for helpful and informative discussions.

\end{acknowledgements}

\bibliographystyle{aa}

\begin{appendix}
\onecolumn
\section{Execution block and weather parameters}

\begin{table}[h!]
\begin{center}
  \caption{Observation and weather parameters for the execution blocks (EBs) as part of the scheduling blocks (SBs) of the various long-baseline observations.}
  {\footnotesize
\begin{tabular}{@{}llllrrrrrrr@{}}
\hline
UID & Date  & Time    &  Source  &  Azm.    & Elev.   & PWV   & Wind speed    & Humidity  & Pressure  & Temperature \\
&      & (UTC)   &          &   (deg) &  (deg)  &  (mm)  & (km\,s$^{-1}$) &  (percent)           & (mbar)    &   (K)      \\
\hline
\hline
\multicolumn{11}{c}{\bf{Juno - Band 6}} \\ 
\hline
\hline
\bf{Xbc7} & 2014/10/19 & 09:04:23 & J0750$+$1231 & 32.48 & 49.01 & 1.656 & 7.70$\pm$0.58 & 60.59$\pm$1.40 & 554.37$\pm$0.01 & 265.11$\pm$0.14\\
\bf{Xdae} & 2014/10/19 & 09:58:59 & J0750$+$1231 & 12.42 & 53.88 & 1.459 & 7.05$\pm$0.55 & 64.15$\pm$1.49 & 554.48$\pm$0.02 & 264.58$\pm$0.13\\
\bf{Xf95} & 2014/10/19 & 10:53:57 & J0750$+$1231 & $-$10.55 & 54.09 & 1.374 & 6.04$\pm$0.86 & 64.08$\pm$1.72 & 554.82$\pm$0.02 & 265.89$\pm$0.16\\
\bf{X117c} & 2014/10/19 & 11:57:46 & J0750$+$1231 & $-$33.96 & 48.42 & 1.447 & 6.05$\pm$2.11 & 35.99$\pm$2.52 & 555.20$\pm$0.04 & 269.25$\pm$0.21\\
\bf{X1363} & 2014/10/19 & 12:50:20 & J0750$+$1231 & $-$48.12 & 40.41 & 1.648 & 6.31$\pm$2.14 & 33.53$\pm$2.25 & 555.53$\pm$0.04 & 270.70$\pm$0.26\\
\hline
\hline
\multicolumn{11}{c}{\bf{Mira - Band 3}} \\ 
\hline
\hline
\bf{X5a8} & 2014/10/17 & 03:12:23 & J0241$-$0815 & 71.19 & 54.30 & 2.819 & 6.59$\pm$0.90 & 39.25$\pm$1.32 & 558.91$\pm$0.03 & 272.05$\pm$0.19 \\
\hline
\bf{X5c0} & 2014/10/25 & 01:45:42 & J0241$-$0815 & 80.13 & 41.94 & 3.625 & 4.72$\pm$0.93 & 24.91$\pm$2.11 & 557.79$\pm$0.06 & 272.04$\pm$0.21 \\
\bf{X926} & 2014/10/25 & 03:08:54 & J0241$-$0815 & 64.67 & 60.29 & 2.738 & 6.46$\pm$1.31 & 40.94$\pm$1.99 & 557.66$\pm$0.07 & 272.81$\pm$0.24 \\
\hline
\hline
\multicolumn{11}{c}{\bf{Mira - Band 6}} \\
\hline
\hline
\bf{X423} & 2014/10/29 & 02:06:36 & J0241$-$0815 & 74.17 & 50.79 & 0.932 & 4.93$\pm$1.07 & 38.79$\pm$1.84 & 556.45$\pm$0.01 & 268.77$\pm$0.31\\
\hline
\bf{X137} & 2014/11/01 & 03:20:36 & J0238$+$1636 & 22.74 & 47.51 & 0.605 & 5.61$\pm$0.93 & 17.48$\pm$2.11 & 555.92$\pm$0.03 & 269.00$\pm$0.21\\
\hline
\hline
\multicolumn{11}{c}{\bf{HL Tau - Band 3}} \\ 
\hline
\hline
\bf{Xc7f} & 2014/10/14 & 06:40:49 & J0510$+$1800 & 29.44 & 43.82 & 1.325 & 4.49$\pm$1.79 & 10.19$\pm$1.55 & 558.64$\pm$0.10 & 274.67$\pm$0.07\\
\bf{Xfd2} & 2014/10/14 & 08:06:58 & J0510$+$1800 & 0.434 & 49.08 & 1.162 & 4.66$\pm$0.96 & 19.32$\pm$1.77 & 558.48$\pm$0.14 & 272.93$\pm$0.12\\
\hline
\bf{X3b2} & 2014/10/28 & 05:31:37 & J0510$+$1800 & 33.42 & 42.12 & 2.805 & 1.46$\pm$0.36 & 22.13$\pm$1.56 & 556.34$\pm$0.02 & 272.81$\pm$0.03\\
\bf{X5fa} & 2014/10/28 & 07:08:10 & J0510$+$1800 & 1.808 & 49.06 & 2.427 & 1.32$\pm$0.62 & 25.21$\pm$1.68 & 555.81$\pm$0.01 & 271.80$\pm$0.08\\
\bf{X845} & 2014/10/28 & 08:43:04 & J0423$-$0120 & $-$61.92 & 50.35 & 1.912 & 3.16$\pm$0.42 & 27.91$\pm$1.70 & 555.69$\pm$0.02 & 271.29$\pm$0.03\\
\hline
\bf{X1f4} & 2014/11/02 & 03:43:20 & J0237$+$2848 & 10.62 & 37.42 & 1.162 & 3.59$\pm$0.90 & 11.25$\pm$2.06 & 556.30$\pm$0.01 & 270.84$\pm$0.08\\
\bf{X444} & 2014/11/02 & 04:56:36 & J0510$+$1800 & 37.44 & 40.06 & 1.070 & 3.92$\pm$0.99 & 10.29$\pm$2.02 & 555.88$\pm$0.02 & 271.09$\pm$0.23\\
\bf{Xc} & 2014/11/02 & 07:58:00 & J0510$+$1800 & $-$22.47 & 46.13 & 0.922 & 5.58$\pm$1.46 & 10.15$\pm$2.06 & 555.10$\pm$0.02 & 270.36$\pm$0.15\\
\hline
\bf{X220} & 2014/11/11 & 02:59:02 & J0423$-$0120 & 65.27 & 47.13 & 2.767 & 7.20$\pm$0.47 & 32.20$\pm$1.93 & 556.89$\pm$0.02 & 269.54$\pm$0.11\\
\bf{X5b2} & 2014/11/11 & 03:54:26 & J0510$+$1800 & 43.74 & 36.03 & 2.561 & 5.48$\pm$1.39 & 30.61$\pm$1.97 & 556.69$\pm$0.02 & 269.03$\pm$0.25\\
\hline
\bf{X22f} & 2014/11/13 & 05:04:53 & J0510$+$1800 & 22.73 & 46.06 & 0.829 & 5.10$\pm$1.00 & 47.84$\pm$1.81 & 553.89$\pm$0.02 & 265.01$\pm$0.24\\
\bf{X461} & 2014/11/13 & 06:11:00 & J0510$+$1800 & $-$0.28 & 49.08 & 0.743 & 6.37$\pm$1.03 & 22.75$\pm$1.80 & 553.37$\pm$0.01 & 266.11$\pm$0.13\\
\bf{X693} & 2014/11/13 & 07:17:19 & J0423$-$0120 & $-$56.14 & 54.88 & 0.811 & 4.67$\pm$1.16 & 18.75$\pm$1.83 & 553.16$\pm$0.01 & 265.32$\pm$0.15\\
\hline
\bf{X306} & 2014/11/14 & 03:49:53 & J0423$-$0120 & 48.49 & 59.31 & 2.024 & 2.38$\pm$0.31 & 73.52$\pm$1.37 & 554.66$\pm$0.01 & 266.36$\pm$0.12\\
\hline
\hline
\multicolumn{11}{c}{\bf{HL Tau - Band 6}} \\ 
\hline
\hline
\bf{Xb2} &  2014/10/24  & 06:48:37 & J0510$+$1800  & 14.37 & 47.92  &  0.678  & 2.33$\pm$0.50  & 29.84$\pm$1.60 & 556.27$\pm$0.02 & 267.42$\pm$0.12 \\
\hline
\bf{Xc8c} & 2014/10/25  & 04:53:59 & J0510$+$1800  & 45.29 & 34.56  &  0.976  & 7.46$\pm$1.03  & 47.70$\pm$1.97 & 557.16$\pm$0.03 & 271.33$\pm$0.17 \\
\hline
\bf{Xdc} &  2014/10/27  & 07:57:13 & J0423$-$0120  & $-$47.52 & 59.78  &  1.351  & 2.92$\pm$0.93  & 18.86$\pm$1.76 & 555.89$\pm$0.01 & 272.00$\pm$0.05 \\
\bf{X33b} & 2014/10/27  & 09:02:35 & J0423$-$0120  & $-$65.28 & 47.12  &  1.198  & 3.29$\pm$0.81  & 16.88$\pm$1.86 & 555.83$\pm$0.03 & 271.49$\pm$0.16 \\
\hline
\bf{X760} & 2014/10/29  & 03:43:49 & J0238$+$1636  & 19.79 & 48.25  &  0.637  & 5.59$\pm$1.07  & 34.65$\pm$1.14 & 556.10$\pm$0.02 & 267.61$\pm$0.12 \\
\bf{X9dd} & 2014/10/29  & 04:53:24 & J0423$-$0120  & 48.26 & 59.43  &  0.445  & 3.93$\pm$1.15  & 29.43$\pm$1.63 & 555.83$\pm$0.02 & 268.10$\pm$0.16 \\
\bf{Xc3c} & 2014/10/29  & 06:02:20 & J0510$+$1800  & 23.21 & 45.93  &  0.333  & 2.95$\pm$0.80  & 25.98$\pm$1.06 & 555.40$\pm$0.03 & 268.38$\pm$0.09 \\
\bf{Xe9b} & 2014/10/29  & 07:09:12 & J0510$+$1800  & $-$0.02 & 49.09  &  0.316  & 3.15$\pm$0.95  & 8.78$\pm$1.43 & 554.95$\pm$0.01 & 270.00$\pm$0.07 \\
\hline
\bf{X387} & 2014/10/31  & 07:35:53 & J0510$+$1800  & $-$12.42 & 48.21  &  0.367  & 5.48$\pm$1.09  & 6.79$\pm$1.70 & 554.09$\pm$0.01 & 268.38$\pm$0.18 \\
\hline
\hline
\multicolumn{11}{c}{\bf{HL Tau - Band 7}} \\ 
\hline
\hline
\bf{X20a} & 2014/10/30 & 04:14:34 & J0423$-$0120 & 58.82 & 52.92 & 0.552 & 3.59$\pm$0.43 & 30.95$\pm$1.92 & 554.86$\pm$0.02 & 268.15$\pm$0.04\\
\bf{X6d8} & 2014/10/30 & 05:41:14 & J0423$-$0120 & 24.09 & 66.59 & 0.584 & 4.16$\pm$1.07 & 26.58$\pm$1.89 & 554.36$\pm$0.02 & 267.43$\pm$0.16\\
\bf{Xa16} & 2014/10/30 & 06:56:18 & J0510$+$1800 & 3.24 & 49.02 & 0.538 & 4.28$\pm$0.96 & 27.46$\pm$1.89 & 553.82$\pm$0.02 & 267.10$\pm$0.06\\
\hline
\bf{X585} & 2014/11/01 & 05:26:32 & J0423$-$0120 & 27.98 & 65.90 & 0.505 & 4.32$\pm$0.63 & 12.05$\pm$2.01 & 555.39$\pm$0.02 & 269.98$\pm$0.24\\
\bf{X826} & 2014/11/01 & 06:40:45 & J0510$+$1800 & 6.02 & 48.88 & 0.461 & 3.41$\pm$1.13 & 8.729$\pm$1.95 & 555.09$\pm$0.02 & 268.85$\pm$0.37\\
\bf{Xacd} & 2014/11/01 & 07:55:44 & J0510$+$1800 & $-$20.48 & 46.65 & 0.427 & 2.57$\pm$0.58 & 3.697$\pm$1.67 & 555.10$\pm$0.00 & 269.13$\pm$0.13\\
\hline
\bf{X2e6} & 2014/11/04 & 03:32:36 & J0237$+$2848 & 11.39 & 37.29 & 0.654 & 7.49$\pm$0.35 & 24.69$\pm$2.21 & 556.37$\pm$0.02 & 267.61$\pm$0.09\\
\bf{X5ab} & 2014/11/04 & 04:48:58 & J0423$-$0120 & 40.68 & 62.53 & 0.613 & 6.20$\pm$1.01 & 19.45$\pm$2.18 & 555.59$\pm$0.02 & 267.06$\pm$0.11\\
\hline
\bf{Xab} & 2014/11/06 & 03:22:56 & J0237$+$2848 & 11.88 & 37.21 & 0.452 & 3.71$\pm$0.63 & 8.410$\pm$1.97 & 558.43$\pm$0.01 & 271.42$\pm$0.08\\
\bf{X446} & 2014/11/06 & 04:37:16 & J0510$+$1800 & 38.33 & 39.55 & 0.424 & 3.49$\pm$0.87 & 9.315$\pm$2.00 & 558.09$\pm$0.02 & 270.92$\pm$0.22\\
\hline
\end{tabular}
}
\label{tab1}
\end{center}
\end{table}

\begin{table}[h!]
\begin{center}
{\footnotesize
\begin{tabular}{@{}llllrrrrrrr@{}}
\hline
UID & Date  & Time    &  Source  &  Azm.    & Elev.   & PWV   & Wind speed    & Humidity  & Pressure  & Temperature \\
&       & (UTC)   &          &   (deg) &  (deg)  &  (mm)  & (km\,s$^{-1}$) &  (percent)           & (mbar)    &   (K)      \\
\hline
\hline
\multicolumn{11}{c}{\bf{SDP.81 - Band 4}} \\ 
\hline
\hline
\bf{Xa1e} & 2014/10/21 & 08:44:54 & J0825$+$0309 & 55.18 & 48.77 & 2.551 & 6.85$\pm$1.34 & 59.35$\pm$1.94 & 555.32$\pm$0.11 & 269.24$\pm$0.15\\
\bf{Xc50} & 2014/10/21 & 09:57:19 & J0825$+$0309 & 30.70 & 60.33 & 2.310 & 6.28$\pm$1.44 & 60.21$\pm$1.79 & 555.53$\pm$0.08 & 269.29$\pm$0.12\\
\bf{Xead} & 2014/10/21 & 11:17:50 & J0825$+$0309 & $-$12.42 & 63.45 & 2.780 & 5.53$\pm$1.33 & 57.69$\pm$1.82 & 556.17$\pm$0.09 & 270.21$\pm$0.17\\
\hline
\bf{X5d0} & 2014/10/24 & 09:12:29 & J0825$+$0309 & 43.68 & 55.69 & 1.143 & 2.31$\pm$0.75 & 22.50$\pm$1.89 & 556.21$\pm$0.02 & 268.89$\pm$0.17\\
\hline
\bf{X12c} & 2014/11/03 & 08:31:31 & J0825$+$0309 & 44.22 & 55.44 & 0.602 & 8.48$\pm$1.39 & 12.29$\pm$2.12 & 554.57$\pm$0.02 & 267.69$\pm$0.08\\
\bf{X34f} & 2014/11/03 & 09:34:20 & J0825$+$0309 & 16.89 & 62.97 & 0.702 & 8.21$\pm$1.39 & 6.595$\pm$1.92 & 554.68$\pm$0.01 & 267.56$\pm$0.14\\
\bf{X572} & 2014/11/03 & 10:36:52 & J0825$+$0309 & $-$17.89 & 62.83 & 0.721 & 5.97$\pm$1.35 & 3.960$\pm$2.01 & 555.27$\pm$0.01 & 268.09$\pm$0.07\\
\bf{X847} & 2014/11/03 & 12:03:41 & J0825$+$0309 & $-$52.10 & 50.97 & 0.739 & 7.56$\pm$1.34 & 4.852$\pm$2.28 & 555.87$\pm$0.02 & 271.71$\pm$0.09\\
\bf{Xa47} & 2014/11/03 & 13:19:17 & J0825$+$0309 & $-$67.85 & 35.79 & 0.788 & 9.45$\pm$1.86 & 8.073$\pm$2.33 & 556.14$\pm$0.04 & 273.78$\pm$0.32\\
\hline
\bf{Xfc6} & 2014/11/11 & 07:27:39 & J0825$+$0309 & 53.83 & 49.77 & 2.952 & 5.07$\pm$1.12 & 30.18$\pm$1.87 & 555.50$\pm$0.01 & 269.40$\pm$0.15\\
\bf{X1634} & 2014/11/11 & 11:17:22 & J0825$+$0309 & $-$47.79 & 53.60 & 3.240 & 5.85$\pm$1.64 & 30.33$\pm$2.36 & 555.69$\pm$0.03 & 272.41$\pm$0.19\\
\bf{X1857} & 2014/11/11 & 12:19:46 & J0909$+$0121 & $-$55.40 & 51.38 & 3.129 & 9.94$\pm$1.82 & 24.75$\pm$2.97 & 555.71$\pm$0.06 & 273.60$\pm$0.23\\ 
\hline
\hline
\multicolumn{11}{c}{\bf{SDP.81 - Band 6}} \\ 
\hline
\hline
\bf{X1484} & 2014/10/12 & 09:13:23 & J0825$+$0309 & 56.85 & 47.44 & 0.951 & 2.60$\pm$0.22 & 12.29$\pm$1.45 & 557.19$\pm$0.01 & 272.99$\pm$0.03\\
\bf{X188b} & 2014/10/12 & 11:16:32 & J0825$+$0309 & 8.320 & 63.76 & 0.973 & 2.76$\pm$0.82 & 4.938$\pm$1.63 & 557.92$\pm$0.02 & 275.89$\pm$0.16\\
\hline
\bf{X11d6} & 2014/10/18 & 09:25:17 & J0825$+$0309 & 47.21 & 53.92 & 3.089 & 4.74$\pm$1.04 & 36.15$\pm$1.36 & 555.50$\pm$0.02 & 267.24$\pm$0.04\\
\hline
\bf{X8be}  & 2014/10/24 & 10:46:04 & J0825$+$0309 & $-$1.17 & 64.00 & 1.247 & 5.54$\pm$0.86 & 24.01$\pm$2.09 & 556.47$\pm$0.06 & 268.33$\pm$0.03\\
\hline
\bf{X1716} & 2014/10/25 & 10:09:26 & J0825$+$0309 & 17.06 & 62.94 & 0.529 & 9.93$\pm$1.48 & 30.90$\pm$2.23 & 555.99$\pm$0.03 & 269.01$\pm$0.35\\
\hline
\bf{X43c}  & 2014/11/02 & 10:39:50 & J0825$+$0309 & $-$17.3 & 62.90 & 0.925 & 3.05$\pm$0.86 & 10.97$\pm$2.33 & 555.89$\pm$0.01 & 270.32$\pm$0.05\\
\bf{X65f}  & 2014/11/02 & 11:43:09 & J0825$+$0309 & $-$44.7 & 55.20 & 0.780 & 3.70$\pm$0.96 & 5.226$\pm$2.20 & 556.30$\pm$0.01 & 273.28$\pm$0.17\\
\hline
\bf{X1099} & 2014/11/08 & 12:09:22 & J0825$+$0309 & $-$58.3 & 46.16 & 0.528 & 1.94$\pm$0.64 & 2.972$\pm$2.28 & 557.66$\pm$0.01 & 277.95$\pm$0.09\\
\hline
\bf{X1f23} & 2014/11/09 & 13:10:58 & J0909$+$0121 & $-$64.9 & 42.67 & 3.000 & 6.21$\pm$2.59 & 1.859$\pm$1.92 & 557.31$\pm$0.06 & 276.16$\pm$0.23\\
\hline
\hline
\multicolumn{11}{c}{\bf{SDP.81 - Band 7}} \\ 
\hline
\hline
\bf{Xb6} & 2014/10/30 & 11:10:02 & J0825$+$0309 & $-$26.6 & 61.30 & 0.353 & 2.12$\pm$0.67 & 9.582$\pm$2.04 & 554.03$\pm$0.01 & 270.72$\pm$0.29\\
\bf{X2eb} & 2014/10/30 & 12:33:58 & J0825$+$0309 & $-$55.8 & 48.25 & 0.382 & 3.99$\pm$1.00 & 0.967$\pm$1.91 & 554.40$\pm$0.02 & 273.98$\pm$0.12\\
\bf{X517} & 2014/10/30 & 13:39:54 & J0909$+$0121 & $-$62.9 & 44.83 & 0.579 & 7.15$\pm$1.81 & 4.879$\pm$2.34 & 554.59$\pm$0.03 & 274.00$\pm$0.15\\ 
\hline
\bf{X5e6} & 2014/10/31 & 08:44:44 & J0825$+$0309 & 43.74 & 55.66 & 0.350 & 5.13$\pm$0.70 & 4.221$\pm$1.56 & 554.15$\pm$0.01 & 268.92$\pm$0.06\\
\bf{X812} & 2014/10/31 & 09:47:57 & J0825$+$0309 & 15.92 & 63.08 & 0.338 & 5.68$\pm$1.16 & 1.629$\pm$1.46 & 554.13$\pm$0.01 & 269.12$\pm$0.28\\
\hline
\bf{Xe29} & 2014/11/01 & 09:22:05 & J0825$+$0309 & 27.00 & 61.22 & 0.554 & 3.36$\pm$0.85 & 1.192$\pm$1.47 & 555.16$\pm$0.01 & 270.97$\pm$0.07\\
\bf{X1059} & 2014/11/01 & 10:25:13 & J0825$+$0309 & $-$7.22 & 63.82 & 0.503 & 2.89$\pm$0.61 & 0.064$\pm$1.62 & 555.39$\pm$0.01 & 271.70$\pm$0.07\\
\bf{X1336} & 2014/11/01 & 11:39:00 & J0825$+$0309 & $-$41.9 & 56.48 & 0.650 & 2.74$\pm$0.69 & $-$0.17$\pm$1.74 & 555.91$\pm$0.02 & 273.99$\pm$0.17\\
\bf{X1562} & 2014/11/01 & 12:42:00 & J0825$+$0309 & $-$59.4 & 45.15 & 0.666 & 3.61$\pm$1.16 & $-$0.58$\pm$1.80 & 556.27$\pm$0.03 & 276.10$\pm$0.16\\
\hline
\bf{Xa99} & 2014/11/04 & 07:11:14 & J0808$-$0751 & 73.87 & 50.51 & 0.606 & 4.73$\pm$1.19 & 24.02$\pm$2.28 & 555.05$\pm$0.02 & 264.39$\pm$0.11\\ 
\bf{Xcc5} & 2014/11/04 & 08:13:54 & J0825$+$0309 & 48.60 & 53.14 & 0.639 & 5.76$\pm$1.18 & 20.94$\pm$2.29 & 554.95$\pm$0.01 & 266.28$\pm$0.17\\
\hline
\end{tabular}
}
\tablefoot{The azimuth and elevation are listed for the bandpass sources from which the optimum WVR scaling is derived. The suffix of the full EB unique identification (UID) is given in the first column as this is all that is required to identify each EB in this work. The UTC time is shown here, although throughout the main text CLT is often used, CLT = UTC - 3\,h. The weather parameters are those averaged over the entire observation of the bandpass source. The double horizontal lines separate the different SBs (i.e. the observation datasets), while the horizontal lines between certain EBs separates those observed on different days. The humidity is the relative humidity measure.}
\end{center}
\end{table}

\clearpage
\newpage

\section{WVR scaling factor}
\begin{table}[h!]
\begin{center}
\caption{Optimal scaling value average over all baselines including the reference antenna for each EB. The improvement in the TPD statistic and also coherence are indicated after applying either the 6 or 12 second optimised scaling value.} 
\begin{tabular}{@{}lrrrrrrr@{}}
\hline
UID &  6 sec. $\phi_{\sigma}$ & 12 sec. $\phi_{\sigma}$ &  32 sec. $\phi_{\sigma}$ &  64 sec. $\phi_{\sigma}$ & TPD   &    \multicolumn{2}{c}{Coherence imp.}     \\
      &                      &                       &                          &                         & imp.   & (60s) & (full)     \\
\hline
\hline
\multicolumn{8}{c}{\bf{Juno - Band 6}} \\ 
\hline
\hline
\bf{Xbc7}$^{*}$  & 1.05$\pm$0.04 & 1.10$\pm$0.08 & 1.20$\pm$0.23 & 1.15$\pm$0.65 & 1.03&   1.00   & 1.01       \\
\bf{Xdae}$^{**}$  & 1.02$\pm$0.08 & 1.04$\pm$0.10 & 1.05$\pm$0.23 & 1.20$\pm$0.45 & 1.00&  1.00   & 1.00    \\
\bf{Xf95}$^{*}$  & 1.08$\pm$0.08 & 1.22$\pm$0.16 & 1.49$\pm$0.30 & 1.71$\pm$0.58 & 1.06&   1.00   & 1.05    \\
\bf{X117c}$^{*}$  & 0.96$\pm$0.04 & 0.96$\pm$0.05 & 0.97$\pm$0.06 & 0.98$\pm$0.14 & 1.03&  1.00   & 1.01     \\
\bf{X1363}$^{**}$  & 1.01$\pm$0.06 & 1.00$\pm$0.05 & 0.96$\pm$0.08 & 0.94$\pm$0.13 & 1.00& 1.00   & 1.00     \\
\hline
\hline
\multicolumn{8}{c}{\bf{Mira - Band 3}} \\ 
\hline
\hline
\bf{X5a8}$^{*}$  & 1.01 $\pm$ 0.05 & 1.16 $\pm$ 0.06 & 1.14 $\pm$ 0.12 & 1.14 $\pm$ 0.31 & 1.13 & 1.00  & 1.00\\
\hline
\bf{X5c0}      & 0.85 $\pm$ 0.12 & 0.99 $\pm$ 0.15 & 0.99 $\pm$ 0.22 & 1.06 $\pm$ 0.27 & 1.05  & 1.00   & 0.98\\
\bf{X926}      & 0.85 $\pm$ 0.10 & 0.93 $\pm$ 0.07 & 0.99 $\pm$ 0.08 & 0.95 $\pm$ 0.27 & 1.09  & 1.00   & 0.99\\
\hline
\hline
\multicolumn{8}{c}{\bf{Mira - Band 6}} \\ 
\hline
\hline
\bf{X423}$^{*}$  & 1.25 $\pm$ 0.12 & 1.30 $\pm$ 0.16 & 1.27 $\pm$ 0.23 & 1.21 $\pm$ 0.26 & 1.07  & 1.01   & 1.07  \\
\hline
\bf{X137}$^{*}$  & 1.15 $\pm$ 0.20 & 1.41 $\pm$ 0.17 & 1.48 $\pm$ 0.37 & 1.44 $\pm$ 0.54 & 1.05  & 1.00   & 1.01  \\
\hline
\hline
\multicolumn{8}{c}{\bf{HL Tau - Band 3}} \\ 
\hline
\hline
\bf{Xc7f}  & 0.68 $\pm$ 0.07 & 0.80 $\pm$ 0.10 & 0.81 $\pm$ 0.24 & 0.79 $\pm$ 0.57 & 1.24   & 1.00   & 1.01\\
\bf{Xfd2}  & 0.91 $\pm$ 0.05 & 0.99 $\pm$ 0.04 & 1.00$\pm$ 0.08 &  0.99 $\pm$ 0.18 & 1.05   & 1.00    &  1.00\\
\hline
\bf{X3b2}  & 0.59 $\pm$ 0.10 & 0.92 $\pm$ 0.11 & 1.10 $\pm$ 0.13 & 1.29 $\pm$ 0.52 & 1.19   &  1.00   & 0.99\\
\bf{X5fa}  & 0.74 $\pm$ 0.07 & 0.93 $\pm$ 0.08 & 1.00 $\pm$ 0.11 & 1.01 $\pm$ 0.15 & 1.16   & 1.00    & 0.99\\
\bf{X845}  & 0.79 $\pm$ 0.08 & 0.91 $\pm$ 0.09 & 0.95 $\pm$ 0.16 & 0.94 $\pm$ 0.26 & 1.12   & 1.00    &  1.00 \\
\hline
\bf{X1f4}  & 0.92 $\pm$ 0.07 & 0.98 $\pm$ 0.07 & 1.00 $\pm$ 0.10 & 0.95 $\pm$ 0.15 & 1.03   &  1.00   & 0.99 \\
\bf{X444}  & 0.99 $\pm$ 0.07 & 1.03 $\pm$ 0.08 & 0.98 $\pm$ 0.13 & 0.91 $\pm$ 0.28 & 1.00    &  1.00   &  1.00\\
\bf{Xc}  & 0.94 $\pm$ 0.07 & 1.04 $\pm$ 0.12 & 1.02 $\pm$ 0.39 & 0.85 $\pm$ 0.65 & 1.01      &  1.00   &  1.00\\
\hline
\bf{X220}  & 0.73 $\pm$ 0.09 & 0.90 $\pm$ 0.12 & 1.03 $\pm$ 0.21 & 1.07 $\pm$ 0.47 & 1.10       &  1.00   & 0.99\\
\bf{X5b2} & 0.73 $\pm$ 0.07 & 0.84 $\pm$ 0.10 & 0.83 $\pm$ 0.19 & 0.71 $\pm$ 0.47 & 1.12       &  1.00   & 1.00 \\
\hline
\bf{X22f}  & 0.93 $\pm$ 0.15 & 1.12 $\pm$ 0.22 & 1.10 $\pm$ 0.34 & 1.00 $\pm$ 0.50 & 1.01       &  1.00   &  1.00\\
\bf{X461}$^{*}$  & 1.03 $\pm$ 0.08 & 1.11 $\pm$ 0.09 & 1.09 $\pm$ 0.16 & 1.06 $\pm$ 0.64 &  1.03 &  1.00   &  1.00\\
\bf{X693}  & 0.92 $\pm$ 0.12 & 1.05 $\pm$ 0.16 & 1.03 $\pm$ 0.29 & 0.74 $\pm$ 0.48 & 1.01        &  1.00   &  1.00\\
\hline
\bf{X306}  & 0.99 $\pm$ 0.11 & 1.07 $\pm$ 0.25 & 1.00 $\pm$ 0.54 & 0.92 $\pm$ 0.82 & 1.00         &  1.00    &  1.00\\
\hline
\hline
\multicolumn{8}{c}{\bf{HL Tau - Band 6}} \\  
\hline
\hline
\bf{Xb2}$^{*}$ & 1.51 $\pm$ 0.18 & 1.49 $\pm$ 0.23 & 1.36 $\pm$ 0.40 & 1.41 $\pm$ 0.46 & 1.12     & 1.01   & 1.03\\
\hline
\bf{Xc8c}$^{**}$   & 0.93 $\pm$ 0.08 & 0.94 $\pm$ 0.09 & 0.96 $\pm$ 0.12 & 0.93 $\pm$ 0.20 &  1.03 &   1.00  & 1.10\\
\hline
\bf{Xdc}$^{*}$  & 1.10 $\pm$ 0.08 & 1.12 $\pm$ 0.12 & 1.12 $\pm$ 0.16 & 1.12 $\pm$ 0.19 &   1.03 &   1.00  & 1.01\\
\bf{X33b}$^{**}$ & 0.98 $\pm$ 0.11 & 0.96 $\pm$ 0.16 & 0.94 $\pm$ 0.22 & 0.95 $\pm$ 0.47 &  1.00 &    1.00 &  1.00\\
\hline
\bf{X760}        & 1.42 $\pm$ 0.10 & 1.40 $\pm$ 0.12 & 1.45 $\pm$ 0.25 & 1.69 $\pm$ 0.44 & 1.25  &   1.02 & 1.08\\
\bf{X9dd}$^{*}$ & 1.49 $\pm$ 0.15 & 1.64 $\pm$ 0.23 & 1.84 $\pm$ 0.67 & 1.63 $\pm$ 1.08 &   1.11 &   1.01 & 1.08\\
\bf{Xc3c}       & 1.35 $\pm$ 0.14 & 1.29 $\pm$ 0.25 & 1.00 $\pm$ 0.60 & 1.06 $\pm$ 0.84 & 1.16   &    1.00 & 1.01\\
\bf{Xe9b}$^{*}$  & 1.38 $\pm$ 0.59 & 1.74 $\pm$ 0.66 & 1.91 $\pm$ 0.85 & 1.64 $\pm$ 1.05  & 1.03 &    1.00 &  1.00\\
\hline
\bf{X387}$^{*}$  & 1.28 $\pm$ 0.41 & 1.66 $\pm$ 0.42 & 2.28 $\pm$ 0.53 & 1.68 $\pm$ 1.06 &  1.03  &    1.00 & 1.01\\
\hline
\hline
\multicolumn{8}{c}{\bf{HL Tau - Band 7}} \\ 
\hline
\hline
\bf{X20a}$^{*}$ & 1.29 $\pm$ 0.22 & 1.52 $\pm$ 0.29 & 1.67 $\pm$ 0.36 & 1.78 $\pm$ 0.72 &  1.12 & 1.01   & 1.07\\
\bf{X6d8}$^{*}$  & 1.28 $\pm$ 0.30 & 1.48 $\pm$ 0.35 & 1.73 $\pm$ 0.45 & 1.65 $\pm$ 0.56 &  1.06 &  1.00  & 1.08 \\
\bf{Xa16}$^{*}$  & 1.24 $\pm$ 0.34 & 1.35 $\pm$ 0.48 & 1.61 $\pm$ 0.75 & 1.83 $\pm$ 0.79 &  1.02 & 1.00   & 1.05\\
\hline
\bf{X585}       & 1.25 $\pm$ 0.20 & 1.27 $\pm$ 0.25 & 1.32 $\pm$ 0.51 & 1.55 $\pm$ 0.93 &  1.03 & 1.00   & 1.01\\
\bf{X826}$^{*}$  & 1.50 $\pm$ 0.23 & 1.60 $\pm$ 0.37 & 1.72 $\pm$ 0.62 & 1.63 $\pm$ 0.98 &  1.07 & 1.01   & 1.04\\
\bf{Xacd}$^{**}$  & 1.29 $\pm$ 0.31 & 1.40 $\pm$ 0.47 & 1.70 $\pm$ 0.83 & 1.24 $\pm$ 0.96 &  1.01 & 1.00   & 1.00\\
\hline
\bf{X2e6}$^{**}$  & 1.17 $\pm$ 0.11 & 1.31 $\pm$ 0.18 & 1.25 $\pm$ 0.41 & 1.12 $\pm$ 0.78 &  1.04 &  1.00  & 1.00\\
\bf{X5ab}$^{**}$  & 1.24 $\pm$ 0.14 & 1.26 $\pm$ 0.19 & 1.07 $\pm$ 0.31 & 1.16 $\pm$ 0.66 &  1.03 &  1.00  & 1.00\\
\hline
\bf{Xab}$^{*}$   & 1.26 $\pm$ 0.28 & 1.27 $\pm$ 0.32 & 1.04 $\pm$ 0.49  & 0.46 $\pm$ 0.46 & 1.01 & 1.00   & 0.99\\
\bf{X446}$^{*}$  & 1.47 $\pm$ 0.09 & 1.52 $\pm$ 0.13 & 1.72 $\pm$ 0.27 & 1.83 $\pm$ 0.28 &  1.25 & 1.00   & 1.11\\
\hline
\end{tabular}
\label{tab2}
\end{center}
\end{table}

\begin{table}[h!]
\begin{center}
\begin{tabular}{@{}lrrrrrrr@{}}
\hline
UID &  6 sec. $\phi_{\sigma}$ & 12 sec. $\phi_{\sigma}$ &  32 sec. $\phi_{\sigma}$ &  64 sec. $\phi_{\sigma}$ & TPD   &    \multicolumn{2}{c}{Coherence imp.}     \\
&                      &                       &                          &                         & imp.   & (60s) & (full)     \\
\hline
\hline
\multicolumn{8}{c}{\bf{SDP.81 - Band 4}} \\ 
\hline
\hline
\bf{Xa1e}       & 0.87 $\pm$ 0.08 & 0.93 $\pm$ 0.07 & 0.93 $\pm$ 0.12 & 0.98 $\pm$ 0.30 & 1.10  & 1.00   & 0.99\\
\bf{Xc50}        & 0.86 $\pm$ 0.12 & 0.96 $\pm$ 0.12 & 1.05 $\pm$ 0.14 & 1.25 $\pm$ 0.31 & 1.08   & 1.00   & 0.97\\
\bf{Xead}        & 0.89 $\pm$ 0.09 & 0.99 $\pm$ 0.08 & 1.04 $\pm$ 0.08 & 1.11 $\pm$ 0.25 & 1.06   & 1.00   & 0.98 \\
\hline
\bf{X5d0}$^{**}$  & 0.97 $\pm$ 0.04 & 1.05 $\pm$ 0.05 & 1.11 $\pm$ 0.06 & 1.24 $\pm$ 0.20 & 1.01 &  1.00  &1.00 \\
\hline
\bf{X12c}$^{*}$  & 1.44 $\pm$ 0.28 & 1.63 $\pm$ 0.34 & 1.76 $\pm$ 0.39 & 1.93 $\pm$ 0.61 &  1.10 &  1.00  & 1.01\\
\bf{X34f}$^{**}$  & 1.33 $\pm$ 0.37 & 1.34 $\pm$ 0.41 & 1.09 $\pm$ 0.65 & 0.95 $\pm$ 0.93 &  1.01 &  1.00  & 1.00\\
\bf{X572}       & 1.03 $\pm$ 0.14 & 1.08 $\pm$ 0.25 & 1.16 $\pm$ 0.40 & 1.24 $\pm$ 0.51 &  1.00 & 1.00   & 1.00\\
\bf{X847}        & 0.94 $\pm$ 0.09 & 0.97 $\pm$ 0.09 & 1.02 $\pm$ 0.13 & 1.03 $\pm$ 0.20 & 1.01 &  1.00  & 1.00\\
\bf{Xa47}${*}$   & 1.22 $\pm$ 0.13 & 1.30 $\pm$ 0.17 & 1.45 $\pm$ 0.27 & 1.72 $\pm$ 0.39 &  1.08 & 1.02   & 1.12\\
\hline
\bf{Xfc6}${*}$  & 0.92 $\pm$ 0.02 & 0.91 $\pm$ 0.02 & 0.90 $\pm$ 0.05 & 0.77 $\pm$ 0.45 &  1.09 &  1.00  & 1.00\\
\bf{X1634}       & 0.95 $\pm$ 0.04 & 0.94 $\pm$ 0.05 & 0.94 $\pm$ 0.06 & 0.90 $\pm$ 0.11 & 1.02 &  1.00  & 1.00\\
\bf{X1857}     & 1.01 $\pm$ 0.03 & 1.00 $\pm$ 0.03 & 0.93 $\pm$ 0.06 & 0.85 $\pm$ 0.14 &  1.00 &  1.00  &1.00 \\
\hline
\hline
\multicolumn{8}{c}{\bf{SDP.81 - Band 6}} \\ 
\hline
\hline
\bf{X1484}$^{**}$ & 0.90 $\pm$ 0.13 & 1.21 $\pm$ 0.20 & 1.40 $\pm$ 0.52 & 1.33 $\pm$ 1.05 & 1.01 &  1.00   &1.00 \\
\bf{X188b}$^{*}$ & 0.89 $\pm$ 0.18 & 1.46 $\pm$ 0.33 & 1.96 $\pm$ 0.52 & 2.20 $\pm$ 0.49 &  1.04 &  1.00  & 1.01\\
\hline
\bf{X11d6}       & 0.91 $\pm$ 0.09 & 1.07 $\pm$ 0.13 & 1.13 $\pm$ 0.18 & 1.06 $\pm$ 0.27 & 1.01  & 1.00   & 0.99\\
\hline
\bf{X8be}       & 0.90 $\pm$ 0.14 & 1.12 $\pm$ 0.18 & 1.31 $\pm$ 0.38 & 1.18 $\pm$ 0.78 & 1.01 & 1.00   & 0.99\\   
\hline
\bf{X1716}$^{*}$  & 1.06 $\pm$ 0.07 & 1.11 $\pm$ 0.09 & 1.07 $\pm$ 0.18 & 1.02 $\pm$ 0.40 &  1.02 & 1.00   & 1.01\\
\hline
\bf{X43c}$^{**}$  & 1.13 $\pm$ 0.13 & 1.18 $\pm$ 0.22 & 1.27 $\pm$ 0.57 & 1.55 $\pm$ 0.80 & 1.01 &  1.00  & 1.00\\
\bf{X65f}$^{**}$ & 0.99 $\pm$ 0.19 & 1.11 $\pm$ 0.25 & 1.28 $\pm$ 0.46 & 1.54 $\pm$ 0.68 &  1.00 &  1.00  & 1.00\\
\hline
\bf{X1099}$^{*}$  & 0.87 $\pm$ 0.19 & 0.88 $\pm$ 0.19 & 0.80 $\pm$ 0.28 & 0.75 $\pm$ 0.42 &  1.02 &  1.00  & 1.00\\
\hline
\bf{X1f23}$^{*}$  & 0.71 $\pm$ 0.10 & 0.66 $\pm$ 0.16 & 0.68 $\pm$ 0.18 & 0.74 $\pm$ 0.25 &  1.14 &  1.02  & 1.04\\ 
\hline
\hline
\multicolumn{8}{c}{\bf{SDP.81 - Band 7}} \\ 
\hline
\hline
\bf{Xb6}$^{**}$   & 1.15 $\pm$ 0.31 & 1.42 $\pm$ 0.40 & 1.46 $\pm$ 0.67 & 1.65 $\pm$ 0.68 &  1.03 & 1.00   &1.01 \\
\bf{X2eb}$^{*}$   & 1.13 $\pm$ 0.12 & 1.19 $\pm$ 0.14 & 1.29 $\pm$ 0.22 & 1.20 $\pm$ 0.41 & 1.03 & 1.00   & 1.01\\
\bf{X517}$^{**}$  & 1.04 $\pm$ 0.11 & 1.05 $\pm$ 0.13 & 1.12 $\pm$ 0.13 & 1.12 $\pm$ 0.16 &  1.00 & 1.00   & 1.03\\
\hline
\bf{X5e6}$^{**}$   & 1.29 $\pm$ 0.49 & 1.61 $\pm$ 0.70 & 2.07 $\pm$ 0.68 & 2.01 $\pm$ 0.98 &  1.02 &  1.00  & 1.00\\
\bf{X812}         & 1.11 $\pm$ 0.27 & 1.29 $\pm$ 0.37 & 1.92 $\pm$ 0.59 & 2.14 $\pm$ 0.84 &  1.00 &  1.00  & 1.00\\
\hline
\bf{Xe29}$^{*}$        & 1.59 $\pm$ 0.36 & 1.67 $\pm$ 0.59 & 1.13 $\pm$ 0.92 & 0.88 $\pm$ 1.13 &  1.02 & 1.00   &1.00 \\
\bf{X1059}        & 1.10 $\pm$ 0.06 & 1.12 $\pm$ 0.11 & 1.13 $\pm$ 0.20 & 1.20 $\pm$ 0.48 & 1.03  &  1.00  & 1.01\\
\bf{X1336}        & 1.14 $\pm$ 0.06 & 1.14 $\pm$ 0.09 & 1.07 $\pm$ 0.25 & 1.17 $\pm$ 0.65 & 1.06  &  1.00  & 1.03\\
\bf{X1562}$^{*}$  & 1.15 $\pm$ 0.05 & 1.17 $\pm$ 0.05 & 1.19 $\pm$ 0.17 & 1.20 $\pm$ 0.35 & 1.09 &  1.00  &1.10 \\
\hline
\bf{Xa99}$^{*}$   & 1.33 $\pm$ 0.11 & 1.42 $\pm$ 0.17 & 1.35 $\pm$ 0.22 & 1.41 $\pm$ 0.53 &  1.12 &  1.01  & 1.02\\
\bf{Xcc5}$^{**}$  & 1.07 $\pm$ 0.13 & 1.16 $\pm$ 0.21 & 1.19 $\pm$ 0.36 & 1.28 $\pm$ 0.45 & 1.01 &  1.00  &1.00 \\
\hline
\end{tabular}
\tablefoot{Water vapour radiometre scale factors are listed where the two-point-deviations $\phi_{\sigma}(T)$ for $T$=\,6, 12, 32, and 64 second timescales are minimised for the corrected phase. The horizontal lines between certain EBs separates those observed on the same date, whereas the double lines separate the different data SBs. The improvement ratio is the statistical improvement of the 6 or 12 second two-point-deviation after the application of the respective optimal scaling factor as compared to the standard WVR correction. The default is to use the 6 second scaling factor, however EBs with a `*' use the scaling factor at the 12 second timescale, while those with `**' use the scaling factor according to Table B.2. The coherence improvements are reported after the input of the RMS phase noise established over (an overlapping) 60\,s and over the full observation of the bandpass source as described in the text.}
\end{center}
\end{table}

\clearpage
\newpage

\begin{table}[h!]
\begin{center}
\caption{Optimal scaling value average over all baselines in each EB.}
\begin{tabular}{@{}lrrrrrrr@{}}
\hline
UID &  6 sec. $\phi_{\sigma}$ & 12 sec. $\phi_{\sigma}$ &  32 sec. $\phi_{\sigma}$ &  64 sec. $\phi_{\sigma}$ & TPD   &    \multicolumn{2}{c}{Coherence imp.}     \\
&                      &                       &                          &                         & imp.   & (60s) & (full)     \\
\hline
\hline
\multicolumn{8}{c}{\bf{Juno - Band 6}} \\ 
\hline
\hline
\bf{Xbc7}$^{*}$   & 1.07 $\pm$ 0.06 & 1.10 $\pm$ 0.10 & 1.11 $\pm$ 0.27 & 1.12 $\pm$ 0.50 &  1.03 & 1.00  &  1.01 \\
\bf{Xdae}$^{*}$   & 1.04 $\pm$ 0.11 & 1.09 $\pm$ 0.13 & 1.18 $\pm$ 0.27 & 1.31 $\pm$ 0.57 &  1.02  & 1.00  &  1.02 \\
\bf{Xf95}        & 1.00 $\pm$ 0.10 & 1.06 $\pm$ 0.17 & 1.16 $\pm$ 0.37 & 1.26 $\pm$ 0.60 &  1.00  & 1.00  &  1.00\\
\bf{X117c}       & 0.97 $\pm$ 0.10 & 0.97 $\pm$ 0.12 & 0.98 $\pm$ 0.15 & 1.00 $\pm$ 0.21 & 1.01   & 1.00  & 0.99 \\
\bf{X1363}$^{*}$  & 1.09 $\pm$ 0.10 & 1.10 $\pm$ 0.11 & 1.11 $\pm$ 0.13 & 1.11 $\pm$ 0.17 &  1.06  & 1.02  & 1.15 \\
\hline
\hline
\multicolumn{8}{c}{\bf{Mira - Band 3}} \\ 
\hline
\hline
\bf{X5a8}$^{*}$  & 1.01 $\pm$ 0.06 & 1.12 $\pm$ 0.07 & 1.15 $\pm$ 0.14 & 1.14 $\pm$ 0.36 & 1.08  & 1.00  & 1.00 \\
\hline
\bf{X5c0}  & 0.89 $\pm$ 0.11 & 1.01 $\pm$ 0.13 & 1.10 $\pm$ 0.19 & 1.12 $\pm$ 0.27 & 1.03 & 1.00   & 0.97 \\
\bf{X926}  & 0.88 $\pm$ 0.07 & 0.98 $\pm$ 0.08 & 1.04 $\pm$ 0.10 & 1.03 $\pm$ 0.24 & 1.08 & 1.00    & 0.99 \\
\hline
\hline
\multicolumn{8}{c}{\bf{Mira - Band 6}} \\ 
\hline
\hline
\bf{X423}$^{*}$ & 1.31 $\pm$ 0.12 & 1.37 $\pm$ 0.15 & 1.42 $\pm$ 0.23 & 1.38 $\pm$ 0.33 &  1.11  & 1.02  & 1.19 \\
\hline
\bf{X137}$^{*}$ & 1.23 $\pm$ 0.28 & 1.37 $\pm$ 0.28 & 1.42 $\pm$ 0.53 & 1.36 $\pm$ 0.74 &  1.06  &  1.00 & 1.02 \\
\hline
\hline
\multicolumn{8}{c}{\bf{HL Tau - Band 3}} \\ 
\hline
\hline
\bf{Xc7f} & 0.72 $\pm$ 0.08 & 0.80 $\pm$ 0.13 & 0.85 $\pm$ 0.30 & 0.91 $\pm$ 0.66 & 1.19   & 1.00  &   1.00\\
\bf{Xfd2} & 0.89 $\pm$ 0.07 & 0.97 $\pm$ 0.07 & 1.01 $\pm$ 0.12 & 0.98 $\pm$ 0.20 & 1.08    &  1.00  &   1.00\\
\hline
\bf{X3b2} & 0.60 $\pm$ 0.13 & 0.91 $\pm$ 0.13 & 1.11 $\pm$ 0.29 & 1.24 $\pm$ 0.55 & 1.14   &  1.00  &  0.99\\
\bf{X5fa} & 0.79 $\pm$ 0.09 & 0.93 $\pm$ 0.10 & 1.00 $\pm$ 0.15 & 1.01 $\pm$ 0.32 & 1.10   &  1.00  & 0.99 \\
\bf{X845} & 0.82 $\pm$ 0.10 & 0.95 $\pm$ 0.12 & 0.98 $\pm$ 0.19 & 0.98 $\pm$ 0.35 & 1.08   &  1.00  &  1.00 \\
\hline
\bf{X1f4} & 0.96 $\pm$ 0.09 & 1.02 $\pm$ 0.10 & 1.01 $\pm$ 0.13 & 1.00 $\pm$ 0.21 & 1.00 &  1.00  &  1.00 \\
\bf{X444} & 0.95 $\pm$ 0.10 & 1.02 $\pm$ 0.11 & 1.01 $\pm$ 0.19 & 0.98 $\pm$ 0.39 & 1.01   & 1.00   &  1.00 \\
\bf{Xc}   & 0.92 $\pm$ 0.10 & 0.99 $\pm$ 0.14 & 1.00 $\pm$ 0.35 & 0.89 $\pm$ 0.64 & 1.01   &  1.00  &  1.00 \\
\hline
\bf{X220} & 0.80 $\pm$ 0.13 & 0.99 $\pm$ 0.16 & 1.12 $\pm$ 0.23 & 1.19 $\pm$ 0.52 & 1.05   & 1.00   &  1.00 \\
\bf{X5b2} & 0.79 $\pm$ 0.13 & 0.92 $\pm$ 0.12 & 0.93 $\pm$ 0.20 & 1.02 $\pm$ 0.53 & 1.07   & 1.00   &  1.00 \\
\hline
\bf{X22f} & 0.88 $\pm$ 0.15 & 0.99 $\pm$ 0.23 & 1.06 $\pm$ 0.34 & 1.14 $\pm$ 0.59 & 1.02   &  1.00  &   1.00\\
\bf{X461}$^{*}$ & 0.98 $\pm$ 0.11 & 1.09 $\pm$ 0.15 & 1.21 $\pm$ 0.30 & 1.26 $\pm$ 0.65 & 1.01  &  1.00  & 1.00  \\
\bf{X693} & 1.00 $\pm$ 0.17 & 1.13 $\pm$ 0.20 & 1.24 $\pm$ 0.10 & 1.24 $\pm$ 0.67 & 1.00  & 1.00   &  1.00 \\
\hline
\bf{X306} & 0.91 $\pm$ 0.17 & 1.03 $\pm$ 0.25 & 1.00 $\pm$ 0.44 & 0.99 $\pm$ 0.68 & 1.00  &  1.00  &   1.00\\
\hline
\hline
\multicolumn{8}{c}{\bf{HL Tau - Band 6}} \\ 
\hline
\hline
\bf{Xb2}$^{*}$        & 1.45 $\pm$ 0.18 & 1.45 $\pm$ 0.23 & 1.39 $\pm$ 0.35 & 1.35 $\pm$ 0.41 & 1.14  & 1.01  & 1.02 \\
\hline
\bf{Xc8c}$^{*}$ & 0.93 $\pm$ 0.10 & 0.92 $\pm$ 0.11 & 0.91 $\pm$ 0.14 & 0.89 $\pm$ 0.19 & 1.05  & 1.01  & 1.12 \\
\hline
\bf{Xdc}$^{*}$       & 1.13 $\pm$ 0.09 & 1.15 $\pm$ 0.13 & 1.16 $\pm$ 0.23 & 1.19 $\pm$ 0.32  & 1.05  & 1.00  & 1.01 \\
\bf{X33b}$^{*}$ & 0.87 $\pm$ 0.13 & 0.85 $\pm$ 0.18 & 0.80 $\pm$ 0.33 & 0.85 $\pm$ 0.53   & 1.02  &  1.00 & 1.00 \\
\hline
\bf{X760}$^{*}$  & 1.42 $\pm$ 0.14 & 1.44 $\pm$ 0.18 & 1.46 $\pm$ 0.30 & 1.51 $\pm$ 0.56 & 1.23  &  1.02 &  1.09\\
\bf{X9dd}       & 1.40 $\pm$ 0.25 & 1.43 $\pm$ 0.37 & 1.58 $\pm$ 0.76 & 1.78 $\pm$ 0.95 & 1.04   &  1.00 &  1.04\\
\bf{Xc3c}       & 1.38 $\pm$ 0.18 & 1.35 $\pm$ 0.27 & 1.15 $\pm$ 0.67 & 1.19 $\pm$ 0.90 & 1.13  &  1.01 &  1.02\\
\bf{Xe9b}       & 1.79 $\pm$ 0.47 & 1.82 $\pm$ 0.61 & 1.66 $\pm$ 0.90 & 1.58 $\pm$ 1.05 & 1.05   &  1.00 &  1.00\\
\hline
\bf{X387}$^{*}$       & 1.59 $\pm$ 0.47 & 1.91 $\pm$ 0.52 & 2.06 $\pm$ 0.72 & 1.96 $\pm$ 0.94 & 1.06  & 1.00  & 1.02 \\
\hline
\hline
\multicolumn{8}{c}{\bf{HL Tau - Band 7}} \\ 
\hline
\hline
\bf{X20a}$^{*}$  & 1.45 $\pm$ 0.27 & 1.59 $\pm$ 0.36 & 1.67 $\pm$ 0.57 & 1.65 $\pm$ 0.73  & 1.22  & 1.02  &1.17  \\
\bf{X6d8}$^{*}$  & 1.46 $\pm$ 0.27 & 1.58 $\pm$ 0.40 & 1.69 $\pm$ 0.58 & 1.68 $\pm$ 0.70  & 1.11  & 1.01  & 1.10 \\
\bf{Xa16}$^{*}$  & 1.43 $\pm$ 0.40 & 1.62 $\pm$ 0.55 & 1.82 $\pm$ 0.72 & 1.87 $\pm$ 0.76  & 1.09  & 1.02  & 1.17 \\
\hline
\bf{X585}$^{*}$  & 1.44 $\pm$ 0.27 & 1.48 $\pm$ 0.36 & 1.49 $\pm$ 0.68 & 1.65 $\pm$ 0.95 & 1.10  & 1.01  & 1.04 \\
\bf{X826}  & 1.52 $\pm$ 0.32 & 1.66 $\pm$ 0.47 & 1.83 $\pm$ 0.72 & 1.71 $\pm$ 0.93  & 1.04  &  1.01 & 1.08 \\
\bf{Xacd}$^{*}$  & 1.54 $\pm$ 0.35 & 1.63 $\pm$ 0.54 & 1.63 $\pm$ 0.84 & 1.41 $\pm$ 0.98 & 1.03  & 1.00  &  1.01\\
\hline
\bf{X2e6}$^{*}$  & 1.17 $\pm$ 0.17 & 1.33 $\pm$ 0.23 & 1.32 $\pm$ 0.48 & 1.34 $\pm$ 0.75 & 1.03  &1.00   & 1.01 \\
\bf{X5ab}$^{*}$  & 1.21 $\pm$ 0.15 & 1.26 $\pm$ 0.20 & 1.32 $\pm$ 0.46 & 1.45 $\pm$ 0.82 & 1.04  & 1.00  & 1.01 \\
\hline
\bf{Xab}$^{*}$   & 1.30 $\pm$ 0.34 & 1.38 $\pm$ 0.46 & 1.30 $\pm$ 0.68 & 0.96 $\pm$ 0.82  & 1.03  & 1.00  & 1.00 \\
\bf{X446}$^{*}$  & 1.39 $\pm$ 0.13 & 1.44 $\pm$ 0.20 & 1.56 $\pm$ 0.39 & 1.60 $\pm$ 0.58  & 1.15  &  1.01 & 1.05 \\
\hline
\end{tabular}
\label{tab3}
\end{center}
\end{table}

\begin{table}[h!]
\begin{center}
\begin{tabular}{@{}lrrrrrrr@{}}
\hline
UID &  6 sec. $\phi_{\sigma}$ & 12 sec. $\phi_{\sigma}$ &  32 sec. $\phi_{\sigma}$ &  64 sec. $\phi_{\sigma}$ & TPD   &    \multicolumn{2}{c}{Coherence imp.}     \\
&                      &                       &                          &                         & imp.   & (60s) & (full)     \\
\hline
\hline
\multicolumn{8}{c}{\bf{SDP.81 - Band 4}} \\ 
\hline
\hline
\bf{Xa1e}        & 0.85 $\pm$ 0.08 & 0.94 $\pm$ 0.11 & 0.97 $\pm$ 0.14 & 0.97 $\pm$ 0.25 & 1.14   & 1.00  &  0.99\\
\bf{Xc50}        & 0.89 $\pm$ 0.13 & 0.98 $\pm$ 0.16 & 1.06 $\pm$ 0.21 & 1.13 $\pm$ 0.33 & 1.04   & 1.00  &  0.97\\
\bf{Xead}        & 0.89 $\pm$ 0.11 & 0.98 $\pm$ 0.10 & 1.02 $\pm$ 0.14 & 1.06 $\pm$ 0.24 & 1.06   &  1.00 & 0.97 \\
\hline
\bf{X5d0}        & 0.93 $\pm$ 0.07 & 1.03 $\pm$ 0.09 & 1.11 $\pm$ 0.16 & 1.17 $\pm$ 0.33 & 1.03   & 1.00  & 1.00 \\
\hline
\bf{X12c}$^{*}$  & 1.27 $\pm$ 0.54 & 1.41 $\pm$ 0.44 & 1.51 $\pm$ 0.60 & 1.65 $\pm$ 0.94 & 1.03  & 1.00  & 1.00 \\
\bf{X34f}$^{*}$  & 1.32 $\pm$ 0.43 & 1.44 $\pm$ 0.57 & 1.55 $\pm$ 0.76 & 1.35 $\pm$ 1.00 & 1.02  & 1.00  & 1.00 \\
\bf{X572}$^{*}$  & 1.05 $\pm$ 0.15 & 1.12 $\pm$ 0.22 & 1.23 $\pm$ 0.42 & 1.24 $\pm$ 0.63 & 1.01  & 1.00  & 1.00  \\
\bf{X847}        & 0.92 $\pm$ 0.07 & 0.93 $\pm$ 0.09 & 0.91 $\pm$ 0.15 & 0.83 $\pm$ 0.23 &  1.03  &1.00   &1.00  \\
\bf{Xa47}        & 0.99 $\pm$ 0.17 & 1.01 $\pm$ 0.24 & 1.11 $\pm$ 0.43 & 1.33 $\pm$ 0.63 &  1.00  & 1.00  &1.00  \\
\hline
\bf{Xfc6}        & 0.94 $\pm$ 0.06 & 0.95 $\pm$ 0.07 & 0.98 $\pm$ 0.18 & 1.08 $\pm$ 0.40 &  1.01  & 1.00  & 1.00 \\
\bf{X1634}       & 0.96 $\pm$ 0.05 & 0.96 $\pm$ 0.05 & 0.97 $\pm$ 0.10 & 0.98 $\pm$ 0.20 &  1.01  & 1.00  & 1.00 \\
\bf{X1857}       & 0.99 $\pm$ 0.05 & 1.00 $\pm$ 0.05 & 1.00 $\pm$ 0.08 & 0.99 $\pm$ 0.15 &  1.00  & 1.00  &  1.00\\
\hline
\hline
\multicolumn{8}{c}{\bf{SDP.81 - Band 6}} \\ 
\hline
\hline
\bf{X1484}$^{*}$  & 0.99 $\pm$ 0.19 & 1.23 $\pm$ 0.27 & 1.36 $\pm$ 0.52 & 1.30 $\pm$ 0.79 & 1.03  & 1.00   & 1.00 \\
\bf{X188b}        & 0.85 $\pm$ 0.23 & 1.28 $\pm$ 0.40 & 1.60 $\pm$ 0.66 & 1.58 $\pm$ 0.89  & 1.01  & 1.00  &1.00  \\
\hline
\bf{X11d6}        & 0.93 $\pm$ 0.13 & 1.10 $\pm$ 0.18 & 1.14 $\pm$ 0.26 & 1.09 $\pm$ 0.46  & 1.01   & 1.00  & 0.99 \\
\hline
\bf{X8be}         & 0.92 $\pm$ 0.13 & 1.10 $\pm$ 0.20 & 1.30 $\pm$ 0.40 & 1.45 $\pm$ 0.65 & 1.01   & 1.00  & 0.99 \\
\hline
\bf{X1716}$^{*}$  & 1.09 $\pm$ 0.13 & 1.18 $\pm$ 0.14 & 1.18 $\pm$ 0.21 & 1.08 $\pm$ 0.46 & 1.05  & 1.00  & 1.00 \\
\hline
\bf{X43c}$^{*}$   & 1.21 $\pm$ 0.17 & 1.28 $\pm$ 0.28 & 1.31 $\pm$ 0.60 & 1.30 $\pm$ 0.90  & 1.02  &  1.00 & 1.00 \\
\bf{X65f}$^{*}$   & 1.13 $\pm$ 0.21 & 1.22 $\pm$ 0.30 & 1.39 $\pm$ 0.57 & 1.64 $\pm$ 0.84  & 1.01  & 1.00  & 1.02 \\
\hline
\bf{X1099}$^{*}$ & 1.06 $\pm$ 0.28 & 1.10 $\pm$ 0.32 & 1.12 $\pm$ 0.47 & 1.11 $\pm$ 0.61 & 1.01  &  1.00  & 1.00 \\
\hline
\bf{X1f23}$^{*}$  & 0.71 $\pm$ 0.13 & 0.71 $\pm$ 0.16 & 0.75 $\pm$ 0.98 & 0.87 $\pm$ 0.32  & 1.15  & 1.02  & 1.05 \\
\hline
\hline
\multicolumn{8}{c}{\bf{SDP.81 - Band 7}}  \\ 
\hline
\hline
\bf{Xb6}$^{*}$   & 1.40 $\pm$ 0.42 & 1.60 $\pm$ 0.53 & 1.62 $\pm$ 0.75 & 1.61 $\pm$ 0.88 & 1.07  & 1.00  & 1.02\\
\bf{X2eb}       & 1.05 $\pm$ 0.24 & 1.09 $\pm$ 0.30 & 1.14 $\pm$ 0.42 & 1.14 $\pm$ 0.65 & 1.00  &  1.00 & 1.00 \\
\bf{X517}$^{*}$  & 1.10 $\pm$ 0.12 & 1.12 $\pm$ 0.14 & 1.18 $\pm$ 0.21 & 1.19 $\pm$ 0.30 & 1.03  &  1.01 & 1.09 \\
\hline
\bf{X5e6}$^{*}$  & 1.71 $\pm$ 0.48 & 1.94 $\pm$ 0.61 & 2.19 $\pm$ 0.68 & 2.02 $\pm$ 0.95  & 1.04  & 1.00  & 1.01 \\
\bf{X812}        & 1.15 $\pm$ 0.49 & 1.25 $\pm$ 0.73 & 1.47 $\pm$ 0.93 & 1.49 $\pm$ 1.13  & 1.00  & 1.00  & 1.00 \\
\hline
\bf{Xe29}       & 1.68 $\pm$ 0.48 & 1.72 $\pm$ 0.66 & 1.39 $\pm$ 0.93 & 1.19 $\pm$ 1.12  & 1.01  & 1.00  & 1.00 \\
\bf{X1059}      & 1.09 $\pm$ 0.07 & 1.08 $\pm$ 0.11 & 1.07 $\pm$ 0.27 & 1.17 $\pm$ 0.51 & 1.02  & 1.00  &  1.01 \\
\bf{X1336}      & 1.13 $\pm$ 0.07 & 1.12 $\pm$ 0.10 & 1.08 $\pm$ 0.28 & 1.28 $\pm$ 0.66 & 1.04   & 1.00  & 1.03 \\
\bf{X1562}      & 1.14 $\pm$ 0.06 & 1.11 $\pm$ 0.09 & 1.09 $\pm$ 0.18 & 1.13 $\pm$ 0.33 & 1.06   & 1.00  & 1.03 \\
\hline
\bf{Xa99}$^{*}$  & 1.36 $\pm$ 0.17 & 1.43 $\pm$ 0.22 & 1.43 $\pm$ 0.33 & 1.39 $\pm$ 0.55 &  1.18  & 1.01  & 1.02 \\
\bf{Xcc5}        & 1.16 $\pm$ 0.19 & 1.19 $\pm$ 0.27 & 1.16 $\pm$ 0.41 & 1.26 $\pm$ 0.58 & 1.01  &  1.00 &  1.01\\
\hline
\end{tabular}
\tablefoot{In some cases around 800 baselines can be in the array configuration and hence establishing the scaling factors with all baselines takes much longer, however the scaling factors and improvement estimates are statistically more robust. The values listed are the same as those in Table B.1. As Table B.1., EBs with a `*' report the 12 second timescale scaling factor. Only EBs that are highlighted as `**' in Table B.1. use the scaling values reported in this table because the reference antenna only analysis reported no improvement.}
\end{center}
\end{table}

\clearpage

\section{Bandpass image results}
\begin{table}[h!]
\begin{center}
\caption{Bandpass calibrator image analysis}
\begin{tabular}{@{}lrrrrrrrrrr@{}}
\hline
UID   &   \multicolumn{3}{c}{\bf Dirty image peaks}        &      \multicolumn{7}{c}{\bf Cleaned image statistics}  \\
\hline
& Peak  &  Peak scale & Imp. &  Peak  & Noise  & S/N  &  Peak scale  & Noise scale  & S/N scale & Imp. \\
   & (Jy)  &  (Jy) &  &  (Jy)  & (mJy)  &  &  (Jy)  & (mJy) &  &  \\

\hline
\hline
\multicolumn{11}{c}{\bf{Juno - Band 6}} \\ 
\hline
\hline
\bf{Xbc7} & 0.504  & 0.509  & 1.01 & 0.501 & 1.990 & 252 & 0.505 & 1.969 & 256 & 1.02  \\
\bf{Xdae} & 0.551  & 0.552  & 1.00 & 0.551 & 1.197 & 461 & 0.552 & 1.190 & 464 & 1.01\\
\bf{Xf95} & 0.551  & 0.551  & 1.00 & 0.546 & 1.674 & 326 & 0.546 & 1.674 & 326 & 1.00\\
\bf{X117} & 0.524  & 0.522  & 1.00 & 0.524 & 1.871 & 280 & 0.522 & 1.894 & 275 & 0.98\\  
\bf{X1363} & 0.408 & 0.413  & 1.01 & 0.403 & 2.984 & 135 & 0.408 & 2.905 & 140 & 1.04\\
\hline
\hline
\multicolumn{11}{c}{\bf{Mira - Band 6}} \\ 
\hline
\hline
\bf{X423} & 0.295  & 0.322  & 1.09 & 0.261 & 3.042 & 86 & 0.299 & 2.914 & 102 & 1.19\\
\hline
\bf{X137} & 1.060  & 1.076  & 1.02 & 1.059 & 6.592 & 161 & 1.076 & 6.340 & 169 & 1.06\\
\hline
\hline
\multicolumn{11}{c}{\bf{HL Tau - Band 3}} \\ 
\hline
\hline
\bf{Xc7f} & 1.161 & 1.167  & 1.01 & 1.118 & 3.340 & 335 & 1.127 & 3.175 & 355 & 1.06\\
\hline
\hline
\multicolumn{11}{c}{\bf{HL Tau - Band 6}} \\ 
\hline
\hline
\bf{Xb2} & 0.853 & 0.863 & 1.01 & 0.840 & 2.432 & 346 & 0.846 & 2.407 & 351 & 1.02\\
\hline
\bf{Xc8c} & 0.677  & 0.729 & 1.08 & 0.629 & 6.679 & 94 & 0.686 & 5.859 & 117 & 1.24\\  
\hline
\bf{Xdc} & 0.946  & 0.946  & 1.00 & 0.944 & 1.339 & 705 & 0.945 & 1.325 & 713 & 1.01\\
\hline
\bf{X760} & 0.969 & 1.027 & 1.06 & 0.943 & 4.355 & 217 & 1.013 & 4.161 & 243 & 1.12\\
\bf{X9dd} & 0.672 & 0.727 & 1.08 & 0.638 & 3.473 & 184 & 0.701 & 2.956 & 237 & 1.29\\
\bf{Xc3c} & 0.669  & 0.682 & 1.02 & 0.652 & 3.069 & 213 & 0.664 & 2.976 & 223 & 1.05\\
\hline
\bf{X387} & 0.825  & 0.833  & 1.01 & 0.823 & 1.746 & 472 & 0.832 & 1.647 & 505 & 1.07\\
\hline
\hline
\multicolumn{11}{c}{\bf{HL Tau - Band 7}} \\ 
\hline
\hline
\bf{X20a} & 0.575  & 0.622  & 1.08 & 0.574 & 2.769 & 207 & 0.615 & 2.256 & 272 & 1.32\\
\bf{X6d8} & 0.574 & 0.614  & 1.07 & 0.559 & 2.214 & 252 & 0.601 & 1.844 & 326 & 1.29\\
\bf{Xa16} & 0.536  & 0.554  & 1.03 & 0.524 & 2.368 & 221 & 0.541 & 2.134 & 253 & 1.15\\

\hline
\bf{X585} & 0.626  & 0.639 & 1.02 & 0.598 & 2.708 & 221 & 0.612 & 2.561 & 239 & 1.08\\
\bf{X826} & 0.480  & 0.500 & 1.04 & 0.463 & 3.653 & 127 & 0.484 & 3.458 & 140 & 1.10\\
\bf{Xacd} & 0.658  & 0.660 & 1.00 & 0.651 & 2.069 & 315 & 0.653 & 2.064 & 316 & 1.00\\
\hline
\bf{X2e6} & 0.644 & 0.635 &  0.99 & 0.641 & 3.326 & 193 & 0.633 & 3.456 & 183 & 0.95\\
\bf{X5ab} & 0.701 & 0.709  & 1.01 & 0.693 & 2.156 & 322 & 0.702 & 2.071 & 339 & 1.05\\
\hline
\bf{X446} & 0.553  & 0.595  & 1.08 & 0.532 & 4.828 & 110 & 0.572 & 4.262 & 134 & 1.22\\
\hline
\hline
\multicolumn{11}{c}{\bf{SDP.81 - Band 4}} \\ 
\hline
\hline
\bf{X12c} & 1.001 &  1.007 &  1.01 & 1.001 & 1.453 & 689 & 1.007 & 1.320 & 763 & 1.07\\
\bf{Xa47} & 0.806 &  0.830 &  1.03 & 0.796 & 5.236 & 152 & 0.814 & 4.699 & 173 & 1.14\\
\hline
\hline
\multicolumn{11}{c}{\bf{SDP.81 - Band 6}} \\ 
\hline
\hline
\bf{X188b} & 0.848 &  0.845 & 1.00 & 0.846 & 1.414 & 598 & 0.843 & 1.534 & 549 & 0.92\\ 
\hline
\bf{X1716} & 0.857  & 0.863 &  1.01 & 0.858 & 2.323 & 369 & 0.863 & 2.217 & 389 & 1.05\\
\hline
\bf{X65f} &  0.994 &  1.010 &  1.02 & 0.971 & 3.604 & 269 & 0.991 &  3.352 & 296 & 1.10\\  
\hline
\bf{X1f23} & 0.477  & 0.488  & 1.02 & 0.475 & 2.008 & 237 & 0.488 & 1.821 & 268 & 1.13\\
\hline
\hline
\multicolumn{11}{c}{\bf{SDP.81 - Band 7}} \\ 
\hline
\hline
\bf{Xb6} & 0.709 &  0.720 &  1.01 & 0.706 & 1.679 & 420 & 0.717 & 1.525 & 470 & 1.12\\
\bf{X2eb} & 0.653 &  0.654 &  1.00 & 0.643 & 2.491 & 258 & 0.644 & 2.519 & 255 & 0.99\\
\bf{X517} & 0.300 &  0.308 &  1.03 & 0.299 & 2.029 & 148 & 0.307 & 1.835 & 167 & 1.13\\
\hline
\bf{X5e6} & 0.638 & 0.644 & 1.01 & 0.631 & 2.414 & 261 & 0.637 & 2.369 & 269 & 1.03\\
\hline
\bf{X1059} & 0.666  & 0.671 & 1.01 & 0.651 & 2.466 & 264 & 0.656 & 2.391 & 274 & 1.04\\
\bf{X1336} & 0.526  & 0.539  & 1.03 & 0.506 & 3.349 & 151 & 0.520 & 3.274 & 158 & 1.05\\
\bf{X1562} & 0.616  & 0.631 & 1.02 & 0.595 & 3.747 & 159 & 0.610 & 3.626 & 168 & 1.06\\
\hline
\bf{Xa99} & 0.519  & 0.529 &  1.02 & 0.518 & 1.945 & 266 & 0.525 & 1.859 & 282 & 1.06\\
\bf{Xcc5} & 0.699  & 0.704 &  1.01 & 0.700 & 2.012 & 348 & 0.705 & 1.942 & 363 & 1.04 \\  
\hline
\end{tabular}
\tablefoot{Image peak, noise, S/N (dynamic range), and image improvement are reported for the cleaned images, whereas only the peak fluxes and improvement are reported for the dirty images. The noise in the normal and scaled WVR images are measured over the same regions.}
\label{tab4}
\end{center}
\end{table}

\clearpage
\newpage

\section{Science target image results}

\begin{table}[h!]
\begin{center}
\caption{Peak flux and map RMS from the science target images processed with and without WVR scaling.}
\begin{tabular}{@{}lrrrrrrrr@{}}
\hline
UID &     Separation     & Peak flux        & Image RMS              &  S/N     &  Peak  flux          & Image RMS     & S/N  & Improvement  \\
 &  to BP cal.        & standard        & standard              &       & scaled          & scaled                &  & ratio  \\
    &  ($^{\circ}$) &  (mJy/bm)    &  ($\mu$Jy/bm)    &          &       (mJy/bm)   &   ($\mu$Jy/bm)     &      &    \\
\hline
\hline
\multicolumn{9}{c}{\bf{Juno - Band 6}} \\ 
\hline
\hline
\bf{Xbc7}$-$a & 8.6   & 7.045 & 84.92 & 83.0 & 7.049 & 84.64 & 83.3 & 1.004\\
\bf{Xbc7}$-$b & 8.6   & 6.945 & 81.60 & 85.1 & 6.962 & 81.54 & 85.4 & 1.003\\
\bf{Xdae}$-$a & 8.6   & 6.512 & 87.96 & 74.0 & 6.516 & 87.99 & 74.0 & 1.000\\
\bf{Xdae}$-$b & 8.6   & 6.865 & 88.95 & 77.2 & 6.852 & 88.90 & 77.1 & 0.998\\
\bf{X1363}$-$a$^*$ & 8.6   & 3.434 & 182.18 & 18.8  & 3.470 & 182.24 & 19.0 & 1.010\\
\bf{X1363}$-$b$^*$ & 8.6   & 3.251 & 156.62 & 20.8 & 3.189 & 157.71 & 20.2 & 0.974 \\
\hline
\hline
\multicolumn{9}{c}{\bf{Mira - Band 6}} \\ 
\hline
\hline
\bf{X423} &  7.6  & 44.30 & 395.5 & 112.0 & 48.94 & 290.1 & 168.7 & 1.506  \\
\hline
\bf{X137} & 20.2   & 65.13 & 163.1 & 399.2 & 65.23 & 143.9 & 453.1 & 1.135  \\
\hline
\hline
\multicolumn{9}{c}{\bf{HL Tau - Band 3}} \\ 
\hline
\hline
\bf{Xc7f} & 9.1   & 2.887 & 26.72 & 108.0 & 2.843 & 26.76 & 106.2 & 0.983\\
\hline
\multicolumn{9}{c}{\bf{HL Tau - Band 6}} \\  
\hline
\hline
\bf{Xb2} & 9.1   & 7.485 & 51.05 & 146.6 & 7.693 & 49.73 & 154.7 & 1.055\\
\hline
\bf{Xc8c} &9.1    & 5.171 & 46.39 & 111.4 & 5.307 & 46.28 & 114.6 & 1.028\\
\hline
\bf{Xdc} & 19.7   & 8.250 & 61.60 & 133.9 & 8.259 & 62.78 & 131.5 & 0.982\\
\hline
\bf{X760} &  27.0  & 5.519 & 38.16 & 144.6 & 5.877 & 38.07 & 154.3 & 1.067\\
\bf{X9dd} &  19.7  & 5.661 & 37.11 & 152.5 & 5.933 & 36.16 & 164.0 & 1.076\\
\bf{Xc3c} & 9.1   & 6.261 & 35.72 & 175.2 & 6.366 & 35.92 & 177.1 & 1.011\\
\hline
\bf{X387} & 9.1   & 6.738 & 38.34 & 175.7 & 6.789 & 38.32 & 177.1 & 1.008\\
\hline
\hline
\multicolumn{9}{c}{\bf{HL Tau - Band 7}} \\ 
\hline
\hline
\bf{X20a} & 19.7   & 8.203 & 80.65 & 101.7 & 8.399 & 81.00 & 103.7 & 1.018\\ 
\bf{X6d8} & 19.7   & 8.057 & 83.93 & 96.0 & 8.250 & 83.85 & 98.4 & 1.024\\
\bf{Xa16} & 9.1   & 8.469 & 82.30 & 102.9 & 8.682 & 83.23 & 104.3 & 1.013\\
\hline
\bf{X585} & 19.7   & 8.369 & 82.44 & 101.5 & 8.516 & 82.42 & 103.3 & 1.017\\
\bf{X826} &  9.1  & 8.083 & 80.82 & 100.0 & 8.245 & 80.65 & 102.2 & 1.022\\ 
\hline
\bf{X5ab} & 19.7   & 7.062 & 91.87 & 76.9 & 7.294 & 91.83 & 79.4 & 1.033\\
\hline
\bf{X446} & 9.1   & 8.548 & 91.83 & 93.1 & 8.622 & 92.08 & 93.6 & 1.005\\
\hline
\hline
\multicolumn{9}{c}{\bf{SDP.81 - Band 6}} \\ 
\hline
\hline
\bf{X1716} &  9.7  & 0.199 & 28.52 & 7.0 & 0.210 & 28.58 &  7.3  & 1.050 \\
\hline
\bf{X65f} &  9.7  & 0.295 & 31.07 & 9.5 & 0.293 & 31.09 & 9.4 & 0.998\\
\hline
\bf{X1f23} &  1.7  & 0.386 & 60.54 & 6.4 & 0.389 & 60.41 & 6.5 & 1.009\\
\hline
\hline
\multicolumn{9}{c}{\bf{SDP.81 - Band 7}} \\ 
\hline
\hline
\bf{Xb6} & 9.7   & 0.261 & 35.81 & 7.3 & 0.262 & 35.83 & 7.3 & 1.005\\
\bf{X517} &1.7    & 0.293 & 45.85 & 6.4 & 0.321 & 45.47 & 7.1 & 1.104\\
\hline
\bf{X5e6} & 9.7   & 0.248 & 32.23 & 7.7 & 0.244 & 32.21 & 7.6 & 0.984\\
\hline
\bf{X1059} & 9.7   & 0.271 & 37.37 & 7.3 & 0.271 & 37.38 & 7.3 & 1.000\\
\bf{X1336} & 9.7   & 0.271 & 38.04 & 7.1 & 0.275 & 38.05 & 7.2 & 1.011\\
\bf{X1562} & 9.7   & 0.323 & 37.97 & 8.5 & 0.324 & 37.92 & 8.6 & 1.003\\
\hline
\bf{Xa99} &  16.1  & 0.270 & 48.23 & 5.6 & 0.275 & 48.21 & 5.7 & 1.016\\
\bf{Xcc5} &  16.1  & 0.281 & 46.06 & 6.1 & 0.274 & 46.01 & 6.0 & 0.983\\
\hline
\end{tabular}
\tablefoot{Separation column indicates the angular separation between the bandpass target and the source using the {\sc casa} analysis utilities `aU.angularSeparationOfFields'. The improvement ratio is with respect to the change in S/N. For Juno there are two science target images made, (a) and (b), per EB to account for the rotation of the asteroid. Although both (a) and (b) images are presented separately the average from each EB is used in the main analysis. The ($^*$) indicates that this Juno EB the calibration is worse with or without WVR scaling. The EBs suffix X12c and Xa47 (SDP.81 B4) have very weak emission such that blends with the noise and image parameters cannot be reported. The noise is measured within the same area for both images with and without WVR scaling applied. The horizontal lines separate the different SBs as noted in Table A.1.}
\label{tab5}
\end{center}
\end{table}

\end{appendix}

\end{document}